\pdfoutput=1
\documentclass[a4paper,fleqn,usenatbib,usedcolumn]{mnras}

\usepackage[T1]{fontenc}
\usepackage{ae,aecompl}

\volume{{\rm accepted}}    

\usepackage{graphicx} 
\usepackage{amsmath}  
\usepackage{amssymb}  
\usepackage{mathabx}  
\usepackage{color}

\newcommand{\RH}{R_{\rm H}}
\newcommand{\AU}{{~\rm AU}}
\newcommand{\J}{{\rm J}}

\title[Survival of habitable planets]{Survival of habitable planets in unstable planetary systems}

\author[Carrera et al.]{
Daniel Carrera,\thanks{E-mail: danielc@astro.lu.se}
Melvyn B. Davies,
and Anders Johansen
\\
Lund Observatory, Department of Astronomy and Theoretical Physics, Lund University, Box 43, SE-221 00 Lund, Sweden\\
}

\date{Accepted XXX. Received YYY; in original form ZZZ}

\pubyear{2016}

\begin{document}
\label{firstpage}
\pagerange{\pageref{firstpage}--\pageref{lastpage}}
\maketitle

%
%
\begin{abstract}
Many observed giant planets lie on eccentric orbits. Such orbits could be the result of strong scatterings with other giant planets. The same dynamical instability that produces these scatterings may also cause habitable planets in interior orbits to become ejected, destroyed, or be transported out of the habitable zone. We say that a habitable planet has \textit{resilient habitability} if it is able to avoid ejections and collisions and its orbit remains inside the habitable zone. Here we model the orbital evolution of rocky planets in planetary systems where giant planets become dynamically unstable. We measure the resilience of habitable planets as a function of the observed, present-day masses and orbits of the giant planets. We find that the survival rate of habitable planets depends strongly on the giant planet architecture. Equal-mass planetary systems are far more destructive than systems with giant planets of unequal masses. We also establish a link with observation; we find that giant planets with present-day eccentricities higher than 0.4 almost never have a habitable interior planet. For a giant planet with an present-day eccentricity of 0.2 and semimajor axis of 5 AU orbiting a Sun-like star, 50\% of the orbits in the habitable zone are resilient to the instability. As semimajor axis increases and eccentricity decreases, a higher fraction of habitable planets survive and remain habitable. However, if the habitable planet has rocky siblings, there is a significant risk of rocky planet collisions that would sterilize the planet.
\end{abstract}


%
%

%
%
\section{Introduction}
\label{sec:intro}

With the discovery by ground-based and space-based surveys that planetary systems are common in the Galaxy \citep[e.g.][]{Mayor_2011,Batalha_2013}, there has been growing interest in whether life bearing worlds may also be common. Most known small planets ---those with radii below $2R_\oplus$--- are in close-in orbits, where they are more easily detected by transit or radial velocity techniques. About 13-20\% of main sequence FGK stars have a $0.8-1.25 R_\Earth$ planet with period up to 85 days \citep{Fressin_2013}. While a simple extrapolation of Kepler data suggests that 20\% of Sun-like stars may have Earth-size planets in the habitable zone \citep{Petigura_2013}, detecting these planets remains a challenge.

Several authors have investigated the orbital stability of hypothetical terrestrial planets in the habitable zones of known planetary systems \citep[e.g.][]{Jones_2001,Menou_2003,Barnes_2004,Rivera_2007}. This type of study provides important constraints on where small planets could reside. However, various authors have noted that the dynamical history of the system is also important. Specifically, some orbits that appear stable today may have been unstable in the past, when the giant planets underwent a dynamical instability. \citet{Veras_2006,Veras_2005} were among the first to study the anti-correlation between giant planets and terrestrial planets. They studied terrestrial planet formation in systems of three giant planets and found that giant planet scatterings interfere with the formation of terrestrial planets. \citet{Raymond_2012,Raymond_2011} extended this work by also including an outer planetesimal disk similar to a primitive Kuiper belt. They found a strong correlation between the formation of terrestrial planets and the presence of large debris belts. More recently, \citet{Matsumura_2013} studied the evolution of fully formed terrestrial planets in a planetary system with three Jupiters. They showed that some orbits that appear stable today will be devoid of rocky planets because of a giant planet configuration that existed in the past.

In this paper we argue that the present-day orbits of observed giant planets contain information about the initial conditions and the dynamical history of the planet system and thus the regions of stability of smaller companion planets. Concretely, we estimate a rough probability that a habitable planet will have survived and remained in the habitable zone to the present day, as a function of the present-day orbit of an observed giant exoplanet.

A planetary system is said to be Hill stable when planet orbits are guaranteed to never cross. Orbit crossings lead to planet-planet scatterings that culminate in planet ejections or physical collisions \citep[e.g.][and references therein]{Davies_2014}. \citet{Juric_2008} found that dynamical instabilities between giant planets naturally explain the eccentricity distribution of observed exoplanets. In planet systems where instabilities occur, habitable planets may be destroyed, or may be moved outside the habitable zone. \citet{Gladman_1993} showed that a planetary system with two planets on circular, co-planar orbits will be Hill stable if $\Delta > 2\sqrt{3}$, where $\Delta$ is the semimajor axis separation measured in mutual Hill radii,

\begin{eqnarray}\label{eqn:Delta}
	\Delta &=& \frac{a_2 - a_1}{\RH}, \\
	\RH   &=& \left( \frac{m_1 + m_2}{3 M_\star} \right)^{1/3} \left( \frac{a_1 + a_2}{2} \right),
\end{eqnarray}
where $m_1$, $m_2$, $a_1$, and $a_2$ are the masses and semimajor axes of the two planets. For systems with more than two planets Hill stability is probably not possible, and systems with $\Delta$ up to 10 have been shown to be always unstable, at least for equal-mass planets \citep{Chambers_1996}. The time to a close encounter grows exponentially with $\Delta$. For a fixed $\Delta$, the time to a close encounter depends weakly on the number of planets and the planet masses, at least up to Jupiter-mass planets \citep{Chambers_1996,Faber_2007}.
For ten Jupiter-mass planets, \citet{Faber_2007} found  $t_{\rm close} \sim 27.7^\Delta$. This extremely steep dependence on $\Delta$ means that two systems with similar $\Delta$ values can have very different lifetimes.

At the time of gas dispersal after a few million years of evolution, the inner region of a protoplanetary disc is believed to be populated by planetesimals and planetary embryos up to the mass of Mars \citep{Kokubo_1996,Johansen_2015}. The assembly of terrestrial planets akin to Earth and Venus occurs by consecutive giant impacts between these embryos over the next 100 Myr \citep{Chambers_1998,Raymond_2006}. In the context of habitable rocky planets, we are therefore interested in planet systems that are stable for at least 100 Myr. However, as noted earlier, a planet system that is stable for a 100 Myr has almost the same $\Delta$ as one that is stable for a few Myr. For this reason we chose to focus on the latter group, so that we can conduct more simulations and produce a more thorough study.

In a related work, \citet{Matsumura_2013} studied the fate of 11 test particles in a flat ($I \sim 0.003^\circ$) planetary system with three Jupiters that had orbit crossings after a few hundred years (figure~1 of their paper). After reproducing their results, we extended their work in several ways:

\begin{enumerate}
\item We moved the giant planets outward and increased their mutual separations so that the orbit crossings happen after a few Myr instead of a few hundred years. As noted earlier, this time-scale is more consistent with terrestrial planet formation.

\item We choose all planet and test particle inclinations from a distribution that results in mutual inclinations of 2-3$^\circ$ (section \ref{sec:methods}), which is in line with exoplanet observations \citep{Johansen_2012}.

\item We explore different giant planet architectures. In addition to the three-Jupiter (3J) systems that \citet{Matsumura_2013} studied, we also explore architectures with four giant planets of unequal masses (4G).

\item Finally, we explore how rocky planets fail to behave like test particles. In a system with multiple rocky planets, collisions and dynamical interactions can have a strong impact on survivability

\end{enumerate}

\citet{Levison_2003} has investigated how the giant planet architecture affects the formation of terrestrial planets. The key difference between our work and theirs is that in their scenario any giant planet instability occurred early, before the formation of rocky planets. In our work, we assume that the giant planet instability occurs after the terrestrial planets have been assembled.

We also differ from previous work in that we focus our attention on habitability. We are not only interested in whether a planet is ejected or destroyed, but also on whether a change in its orbital parameters can take the planet out of the habitable zone, or otherwise render it uninhabitable. The habitable zone is the region around a star where a rocky planet with the right atmosphere can have liquid water on the surface.

Traditional estimates of the habitable zone are produced by 1D climate models which assume cloud-free, saturated atmospheres \citep[e.g.][]{Kasting_1993,Selsis_2007,Kopparapu_2013a}. The inner edge of the habitable zone is set by the runaway greenhouse or the moist greenhouse limit. In the latter, the stratosphere becomes water rich, leading to photo-dissociation and the loss of water through hydrogen escape \citep{Kopparapu_2013a}. All 1D models tend to give pessimistic estimates of the inner edge of the habitable zone because they cannot include the cooling effect of cloud feedback (which increases albedo), long-wave emission from subsaturated air above the subtropics, or heat transport away from the equator \citep{Wolf_2014}. For this reason we refer to the 1D models as the ``conservative'' habitable zone.

Despite their limitations, 1D models are commonly used because they are cheaper than full 3D models (GCMs) and can provide habitable zone limits for a wide range of stellar parameters. Besides, there are many other factors that affect habitability. Rocky planets much dryer than Earth (``Dune'' worlds) can avoid the moist greenhouse much closer to their parent star \citep{Abe_2011}, while eccentricity and obliquity can make a planet more resilient against global freezing \citep[e.g.][]{Dressing_2010,Armstrong_2014}. \citet{Williams_2002} used a GCM to study the climate of an Earth-like planet on an eccentric orbit around a Sun-like star. They found that the moist greenhouse limit is set primarily by the mean orbital flux received by planet, which is given by

\begin{equation}
	\langle F \rangle = \frac{L}{4\pi a^2 (1-e^2)^{1/2}},
\end{equation}
where $a$ and $e$ and the planet's semimajor axis and eccentricity, $L$ is the stellar luminosity, and $\langle F \rangle$ is the mean orbital flux.

This paper is organised as follows. In section \ref{sec:methods} we describe our numerical methods and initial conditions. We use an N-body integrator to model the long term evolution of different types of planetary systems. We select orbital separations and inclinations consistent with observation. We model terrestrial planets both as test particles, and as massive bodies. In section \ref{sec:results} we present our results, and in section \ref{sec:discussion} we discuss the implications. Finally, we summarize and conclude in section \ref{sec:conclusions}.

%
%
\section{Methods}
\label{sec:methods}

We performed N-body simulations of planetary systems using the hybrid integrator of the MERCURY code \citep{Chambers_1999}. We simulated systems with either three Jupiter-mass planets (3J) or four giant planets of unequal masses (4G) orbiting a Sun-like star. In all our simulations the planets are initially in circular orbits with the innermost giant at 5 AU and the other giant planets at fixed separations in $\Delta$ (Eqn.~\ref{eqn:Delta}). The planet masses and $\Delta$ values are shown in Table \ref{tab:initial}. The planet system 4Gb has the same giant planet masses as the solar system, 4Ga is less hierarchical than the solar system, and 4Gc is more hierarchical. The value of $\Delta$ was chosen so that the systems would typically have orbit crossings after a few Myr (we discussed the rationale in section \ref{sec:intro}). We ran each system for 30 Myr unless noted otherwise. In all our runs, when a planet or test particle reaches a distance of 1000 AU from the star, it is considered an ejection, and is removed from the simulation.

Following the prescription of \citet{Johansen_2012}, we give each orbit a random inclination $I$ between 0$^\circ$ and 5$^\circ$ and a random longitude of ascending node ($0^\circ < \Omega < 360^\circ$). This results in the planets having having a range of mutual inclinations between 0$^\circ$ and 10$^\circ$ with typical values around 2--3$^\circ$, which is consistent with systems of super-Earths observed with the Kepler telescope \citep{Johansen_2012}. We also give the planets random mean anomalies ($0^\circ < \lambda < 360^\circ$).

\begin{table}
	\centering
	\caption{Initial conditions. We model 3-4 giant planets with the inner giant at 5 AU. The other giant planets are added at fixed separations $\Delta$ (Eqn.\,\ref{eqn:Delta}). The value of $\Delta$ is chosen to produce close encounters in a few million years. The orbits are circular, and the mutual inclinations are typically 2--3$^\circ$ (see main text). See also Fig.~\ref{fig:initial}.}
	\label{tab:initial}
	\begin{tabular}{l|lllccc}
		\hline
		 & $m_1 / M_\J$ & $m_2 / M_\J$ & $m_3 / M_\J$ & $m_4 / M_\J$ & $a_1$/AU & $\Delta$\\
		\hline
		3J  & 1 & 1    & 1    &  -   & 5 & 5.1 \\
		4Ga & 1 & 0.35 & 0.10 & 0.05 & 5 & 5.7 \\
		4Gb & 1 & 0.30 & 0.05 & 0.05 & 5 & 5.7 \\
		4Gc & 1 & 0.20 & 0.05 & 0.05 & 5 & 5.7 \\
		\hline
	\end{tabular}
\end{table}

\begin{figure}
	\includegraphics[width=\columnwidth]{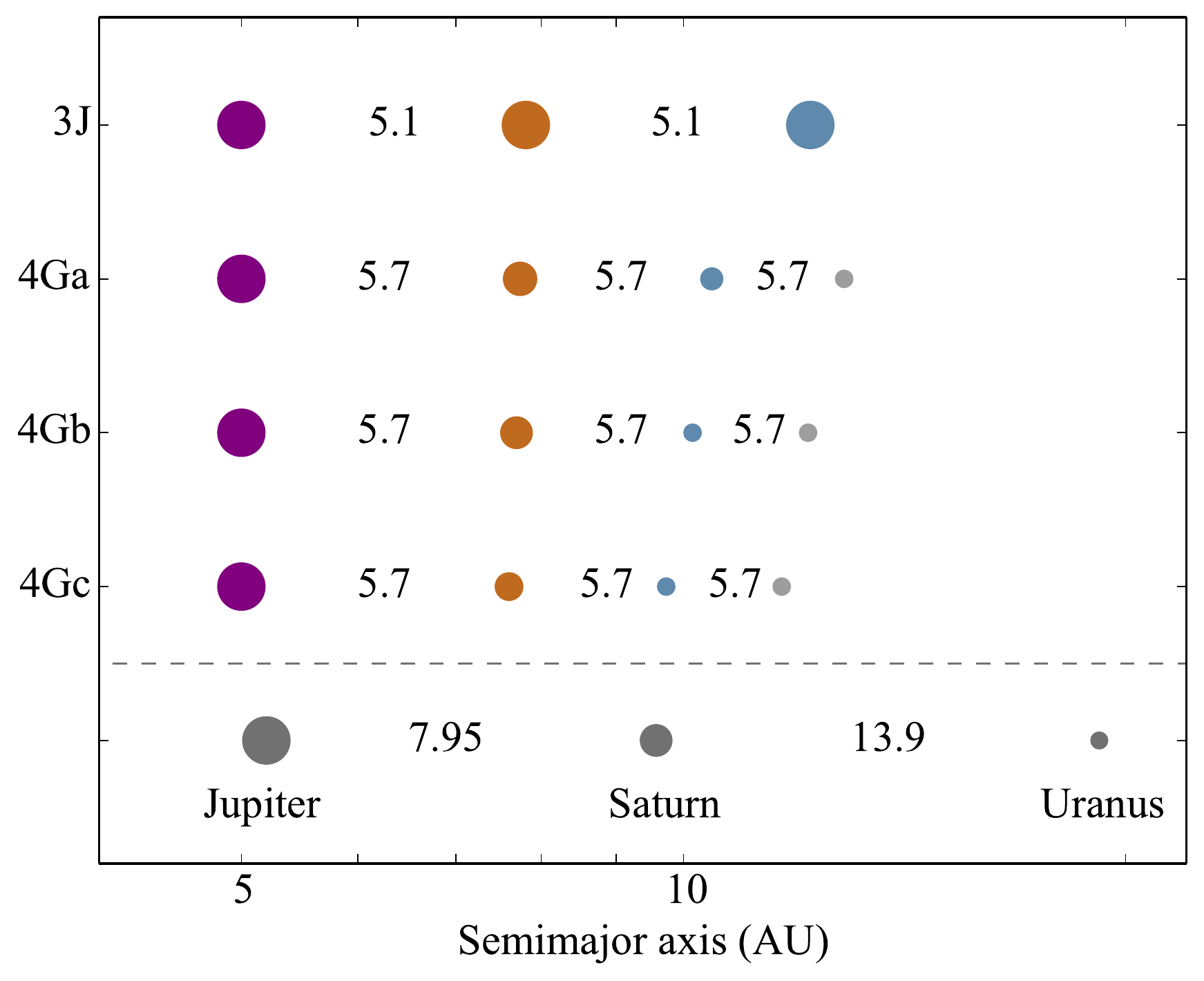} \\
	\includegraphics[width=\columnwidth]{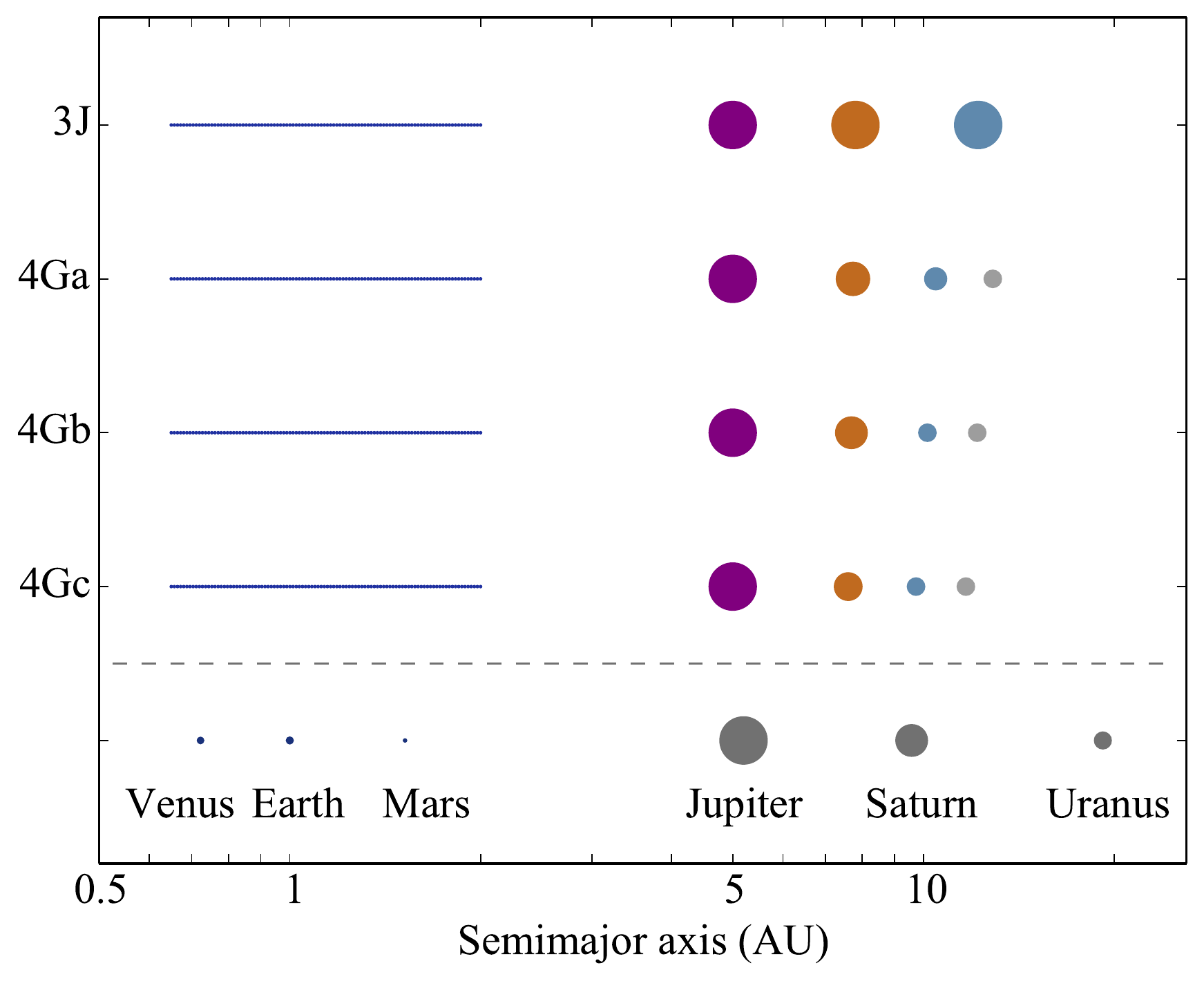}
    \caption{Initial conditions. We consider equal-mass (3J) and hierarchical (4G) giant planet systems. Each planet is represented as a circle with radius proportional to $m^{1/3}$. The top plot shows the separations between planets in terms of mutual Hill radii (Eqn.\,\ref{eqn:Delta}). The bottom plot shows the 100 test particles between 0.65 and 2 AU. Solar system shown for scale.}
    \label{fig:initial}
\end{figure}

In our first set or runs, we ran 50 instances of each planet system in Table \ref{tab:initial}, along with 100 test particles. The test particles have semimajor axes distributed uniformly in log from 0.65 to 2 AU. The other orbital parameters ($e, I, \Omega, \lambda$) follow the same prescription as the massive planets. The initial conditions of the giant planets and test particles are illustrated in Fig.~\ref{fig:initial}. There were six 3J runs that did not experience ejections or collisions within the 30 Myr integration time. We extended these runs for another 30 Myr, at which point four more runs had experienced ejections and collisions. The two remaining runs were prolonged for an additional 30 Myr, but no ejections or collisions occurred. We consider those two runs ``unresolved'' and exclude them from our final results.

In our second set of runs we studied the effect of rocky planet mass and multiplicity. In general, single rocky planet should behave similar to a test particle, but test particles do not capture rocky planet collisions or dynamical interactions (see section \ref{sec:rocky}). We performed a new set of 4Gb simulations with the test particles replaced by 1, 2, or 4 Earth-mass planets as described in Table \ref{tab:initial-rocky}. As before, the other orbital parameters ($e, I, \Omega, \lambda$) follow the prescription of the giant planets. We ran 50 instances of each system. The 1-Earth runs (4Gb+1e) serve as a test for the particle runs, while the other runs show the effect of multiplicity in the terrestrial zone.  We placed the rocky planets in the orbits of the test particles that were closest to present-day Mercury, Venus, Earth, and Mars.

As a final set of runs, we chose the 4Gb and 4Gb+4e runs for a more in-depth study. We ran an additional 250 runs or 4Gb+4e, for a total of 300 runs. We added 40 test particles to 4G, for a total of 140 particles from 0.65 to 3.15 AU, and ran a new set of 300 runs.

\begin{table}
	\centering
	\caption{We performed additional runs with the 4Gb planets, but with the test particles replaced by one, two, or four Earth-mass planets in the terrestrial zone. We use 4Gb+1e to verify the results of the test particle runs. We use 4Gb+2e and 4Gb+4e to study the dynamical interactions between rocky planets. The Earth-mass planets were placed in the orbits of the test particles closest to present-day Mercury, Venus, Earth, and Mars.}
	\label{tab:initial-rocky}
	\begin{tabular}{c|ccccc}
		\hline
		 & $a_1$/AU & $a_2$/AU & $a_3$/AU & $a_4$/AU & masses ($M_\Earth$) \\
		\hline
		4Gb+1e & -    & -    & 1.00 & -    & 1 \\
		4Gb+2e & -    & 0.72 & 1.00 & -    & 1 \\
		4Gb+4e & 0.39 & 0.72 & 1.00 & 1.52 & 1 \\
		\hline
	\end{tabular}
\end{table}

%
%
\section{Results}
\label{sec:results}

\subsection{System architecture}
\label{sec:arch}

\begin{figure*}
	\includegraphics[width=1.4\columnwidth]{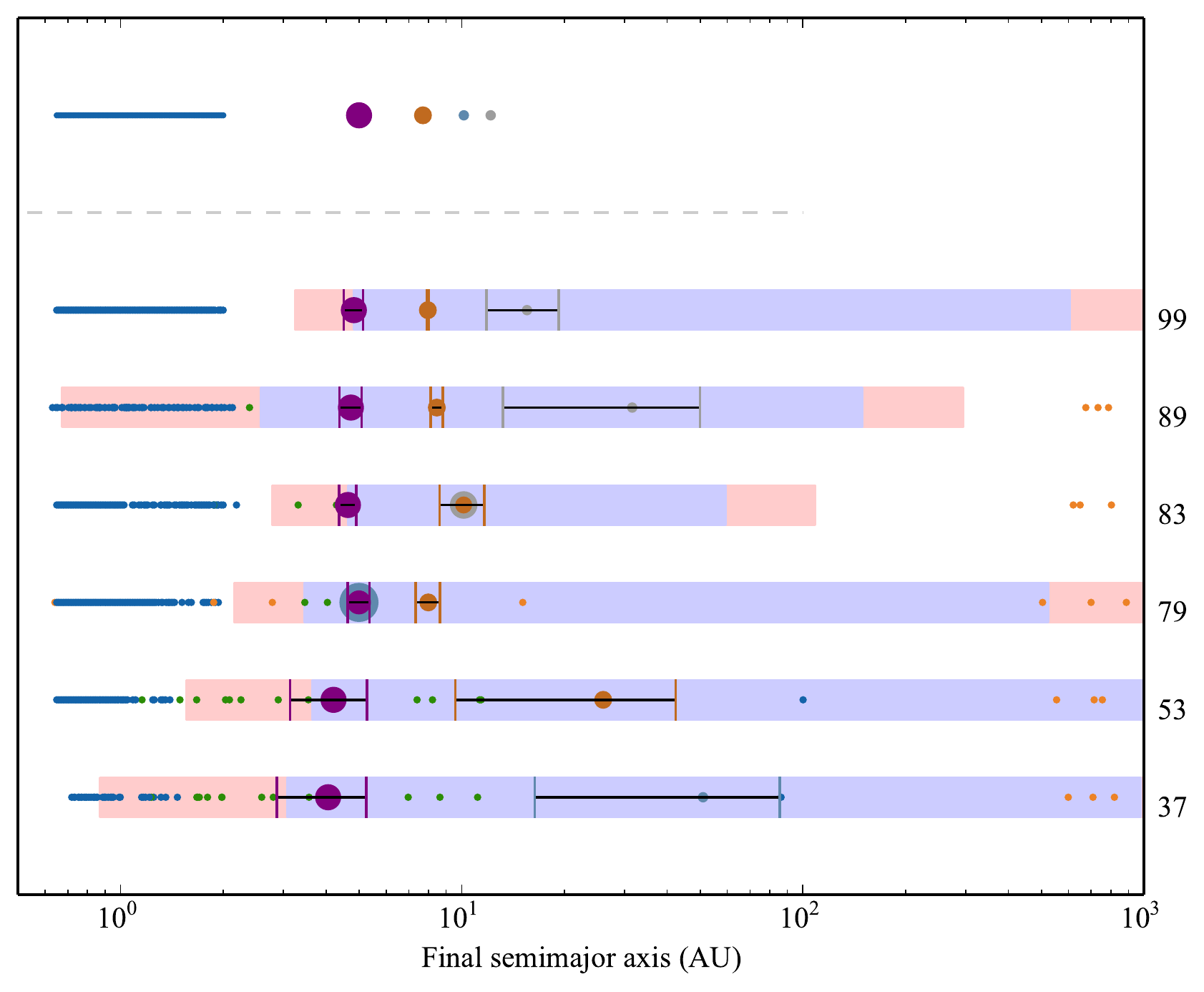}
    \caption{Summary of six typical runs for 4Gb. Each planet is represented as a circle of radius proportional to $m^{1/3}$ drawn at the planet's semimajor axis with a horizontal line from periastron to apastron. Test particles are shown in their last known location with a colour that indicates whether the particle survived (blue), hit the star (green), or was ejected (orange). On the right side we show the number of surviving particles. The initial conditions are shown at the top. When two planets collide (middle two runs) we alter the planet radius to make the colour of the two planets easier to see. The light backgrounds show the range of the giant planet semimajor axis (light blue) or position (light red) over the course of the simulation.}
    \label{fig:lovis_subset}
\end{figure*}

Figure \ref{fig:lovis_subset} shows typical results from our 4Gb runs. Because the system is chaotic, two systems with similar orbits can have radically different histories. \citet{Matsumura_2013} found that giant planet-planet collisions are associated with higher survival rates. Although we can certainly reproduce this result for 3J systems, we found no such correlation for 4G systems. In addition, we found that higher inclinations and wider orbits significantly reduce the number of collisions. Using the same initial conditions as \citet{Matsumura_2013} (all 3J), we found that $\sim 78\%$ of the systems had at least one collision between giant planets, and $\sim 12\%$ had multiple collisions. In contrast, using our initial conditions (section \ref{sec:methods}), only $\sim 10\%$ of the systems experience giant planet collisions, and none of our 3J systems had two collisions. This means that, for wider orbit giant planets, planet-planet collisions have a diminished role in shaping the evolution of the system.

\begin{figure*}
	\includegraphics[width=0.89\textwidth]{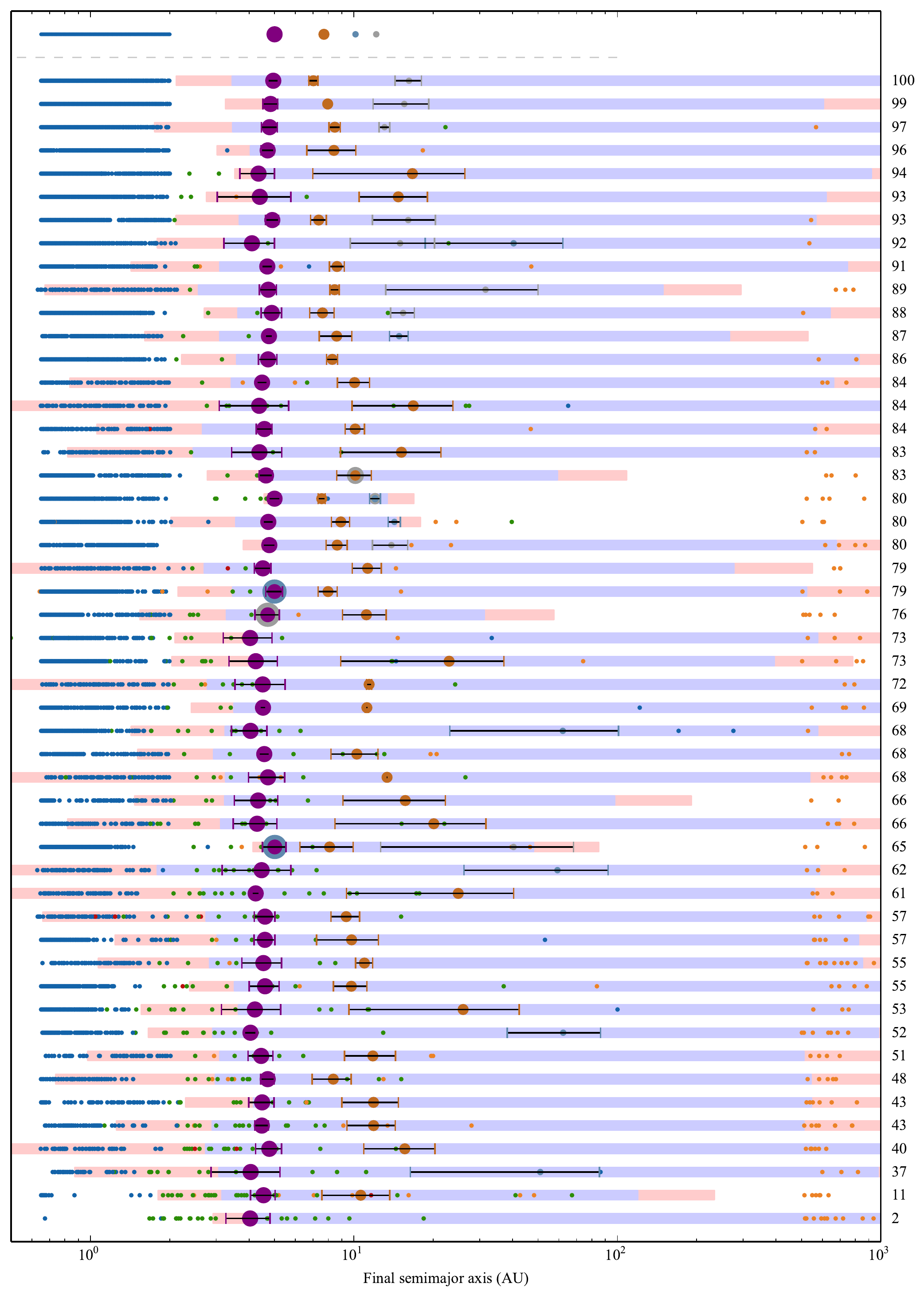}
    \caption{Final result for all 50 runs of 4Gb, using the same symbols as Fig.~\ref{fig:lovis_subset}. In addition, when a test particle collides with a giant planet, we mark its last recorded position in red. 4G systems have a high survival rate with a median of 73\% of test particles surviving for 4Gb. 4G systems that have a collision between giant planets seem to have the same survival rate as the rest of the population (contrast with 3J in Fig.~\ref{fig:lovis_all_3J}).}
    \label{fig:lovis_all_4G}
\end{figure*}

\begin{figure*}
	\includegraphics[width=0.89\textwidth]{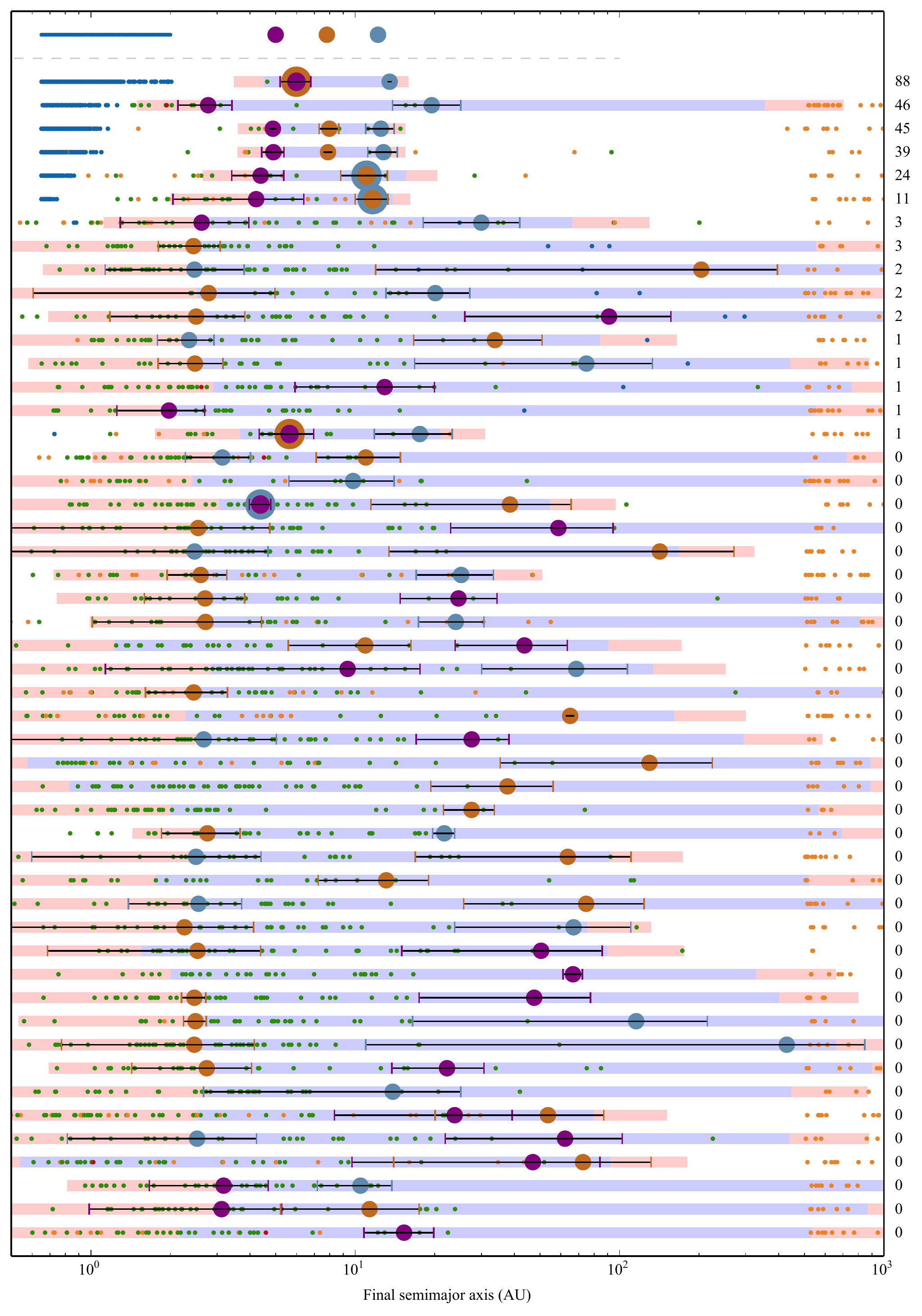}
    \caption{Final result for all 50 runs of 3J, using the same symbols as Fig.~\ref{fig:lovis_all_4G}. There are two runs that still have three giant planets left after 90 Myr of integration. We consider these runs unfinished and exclude them from the final results. All other runs that have surviving particles and two giant planets are Gladman stable ($\Delta > 2\sqrt{3}$). 3J systems are extremely destructive, with a mean survival rate of 3.9 particles (excluding the two unresolved runs) and a median of zero. 3J systems that have giant planet collisions are less destructive, with a mean survival rate of 25 particles and a median of 11.}
    \label{fig:lovis_all_3J}
\end{figure*}

Our most salient result is that hierarchical (4G) planet systems are significantly less destructive to terrestrial planets than equal-mass (3J) systems. A similar result was found by \citet{Veras_2006} for systems with three giant planets. Figures \ref{fig:lovis_all_4G} and \ref{fig:lovis_all_3J} show the full set of results for 4Gb and 3J. Note that test particles usually survive in 4Gb, and they rarely survive in 3J. Out of the four runs that had ejections after a 30 Myr extension, one run had 3 survivors and the others had none. In other words, the runs that took longer to become unstable had the same survival rate as the other runs, within statistical uncertainty. For this reason, excluding the two remaining unresolved runs from the final analysis should give the most unbiased result.

An interesting feature of Fig.\,\ref{fig:lovis_all_4G} is that even when a giant planet wanders well inside the terrestrial zone (e.g.\,light red bar extends beyond the left of the frame) there is little effect on the test particles, with up to 84\% of the particles surviving. Contrast this with Fig.\,\ref{fig:lovis_all_3J}, where there are zero surviving particles inside the light red bar. The key difference is that in the 3J systems every intruder is a Jupiter-mass planet. In the 4Gb runs, we verified that every single intruder is one of the two Neptune-mass planets. This has two important implications,

\begin{itemize}
\item The volume traced by the Hill sphere of a Neptune-mass planet as it enters the terrestrial zone is seven times smaller than that of a Jupiter-mass planet. The incursions typically last for a few thousand years. Close encounters between the test particle and the giant planet are not likely to occur in this short interval of time. Also, with an orbital inclination as low as 1.5$^\circ$, the Hill sphere of a Neptune-mass planet could completely miss the path of a rocky planet.

\item As a general rule, a giant planet can only eject a test particle after a single encounter if planet's escape speed is greater than the local orbital speed \citep[e.g.][]{Goldreich_2004,Ford_2008,Davies_2014}. This means that a single encounter with a Neptune-mass planet cannot eject a particle inside 1.6 AU, whereas a Jupiter-mass planet can eject particles in a single encounter as close as 0.24 AU.
\end{itemize}

\begin{figure}
	\includegraphics[width=\columnwidth]{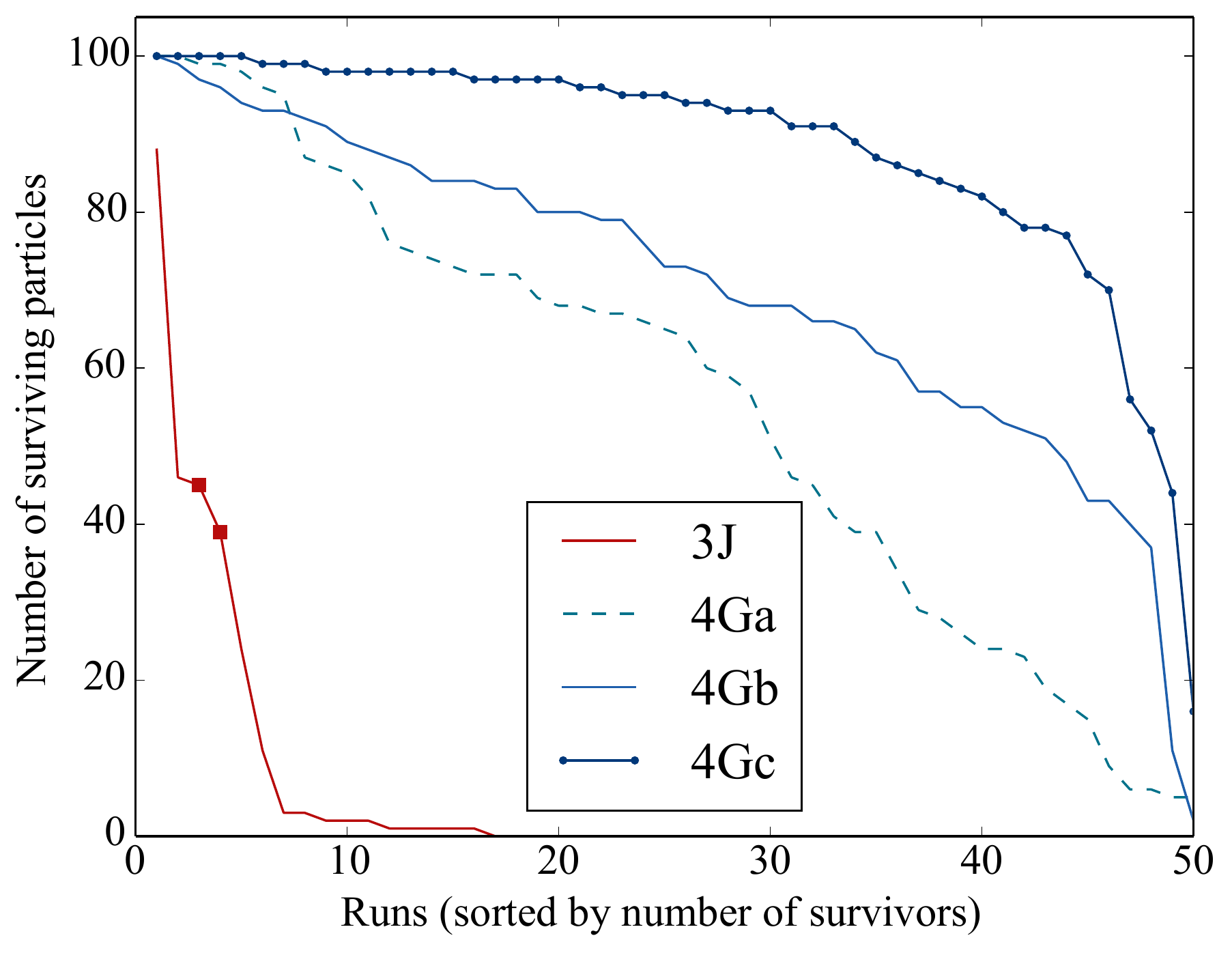}
    \caption{Number of surviving test particles for each run. We did 50 runs for each architecture (3J, 4Ga, 4Gb, 4Gc) and each run had 100 test particles. We sorted the runs from the most survivors (left) to fewest (right). The two 3J runs that are unresolved are included in the plot but marked with red squares. Not only is 3J more destructive than than a hierarchical system (4G), but within 4G, the less hierarchical (e.g.\,4Ga) the more destructive.}
    \label{fig:survivors}
\end{figure}

Figure \ref{fig:survivors} shows the number of survivors for 3J, 4Ga, 4Gb, 4Gc. Not only are the hierarchical (4G) systems much less destructive than the equal-mass (3J) system, but there is a clear correlation among the 4G systems where the more hierarchical systems (e.g.~4Gc) the higher the survival rate of test particles. Table \ref{tab:final-particle} shows the fate of the test particles in more detail. In all cases, approximately two-thirds of the particles lost are ejected from the system, and one-third collide with the Sun. Collisions with the giant planets are very rare.

\begin{table}
	\centering
	\caption{The fate of test particles in the 3J, 4Ga, 4Gb, and 4Gc systems. Particles are lost primarily by being ejected from the system. The ejection criterion is that the particle reaches 1000 AU. Collisions with giant planets are very rare.}
	\label{tab:final-particle}
	\begin{tabular}{l|rrrr|rrr}
		\hline
        & 3J & 4Ga & 4Gb & 4Gc \\
		\hline
        Ejected      & 63.1\% & 32.1\% & 22.3\% &  9.2\% \\
        Hit the Sun  & 32.9\% & 12.0\% &  7.3\% &  2.7\% \\
        Hit a planet &  0.1\% &  0.2\% &  0.2\% &  0.0\% \\
        Survived     &  3.9\% & 55.6\% & 70.2\% & 88.1\% \\
		\hline
	\end{tabular}
\end{table}

\begin{figure}
	\includegraphics[width=\columnwidth]{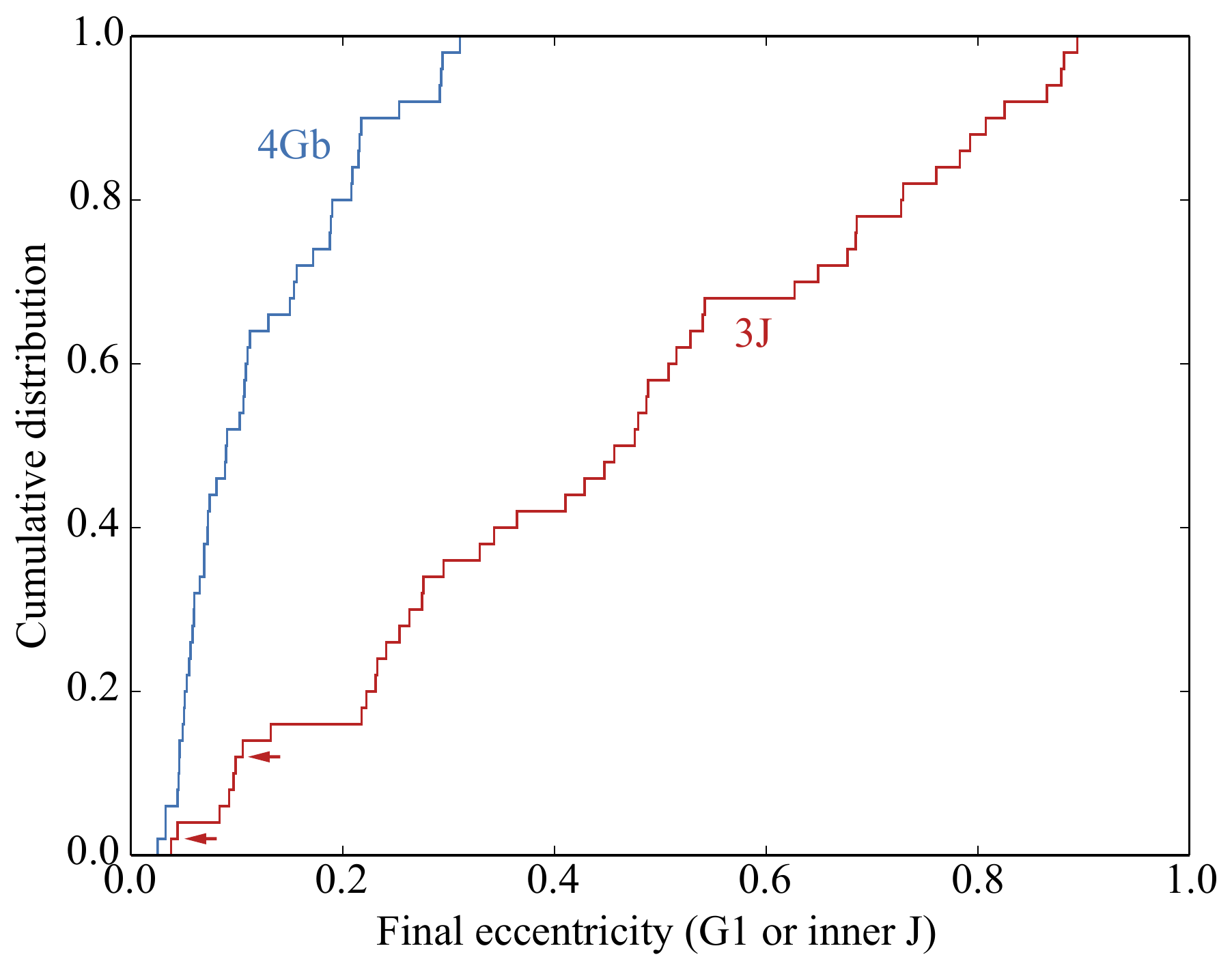}
    \caption{Final giant planet eccentricities after the instability. The two 3J runs that are unresolved are included in the plot but marked with small arrows. Observed giant planets with eccentricities higher than 0.4 most likely came from systems similar to 3J, and it is likely that terrestrial planets never formed, or were destroyed when the instability occurred. Giant planets with $e < 0.2$ are likely to either never have become unstable, or have belonged to a 4G-type system that usually leaves habitable planets intact.}
    \label{fig:eccentricities}
\end{figure}

The next most salient result is that 3J systems leave giant planets in more eccentric orbits, and they have more incursions into the terrestrial zone. A correlation between planet mass ratios and final eccentricity is expected from the conservation of energy and angular momentum. Indeed, a similar correlation was found by \citet{Ford_2008} for two-planet systems. The authors argue that planet-planet scatterings can reproduce the observed distribution of eccentricities from a distribution of planet mass ratios. Figure \ref{fig:eccentricities} shows the cumulative distribution of eccentricities at the end of the run for the 3J and 4Gb runs. As a rule of thumb, if a giant planet has a present-day eccentricity above 0.4, it is very likely that it came from a very equal-mass system, similar to the 3J systems that we have modelled. In turn, giant planets with eccentricity lower than 0.2 probably came from a hierarchical system, or acquired their present-day eccentricity through some other mechanism. For eccentricities between 0.2 and 0.4, it is more difficult to know what the initial system might have been like. The answer probably depends mostly on the giant planet initial-mass function, which is currently unknown \citep[e.g.][]{Juric_2008}.

One might object that the 4G systems have a lower total amount of mass than 3J. We have verified that the difference between 4G and 3J (Fig.~\ref{fig:eccentricities}) is not mainly due to the total mass, but is mainly due to the mass ratios. We performed a set of runs where we doubled the masses of the 4Ga planets (but kept $\Delta$ fixed) and the results were consistent with the original 4Ga runs. In other words, the effect of giant planet mass is secondary. The most important variables are $\Delta$ (which sets the stability time-scale), and the mass ratios. In the case of 3J systems, mutual inclination is also important. We also performed runs with very flat 3J systems ($I \sim 0.003^\circ$, similar to \citet{Matsumura_2013}) and those have a much higher survival rate. This is due in part to a greater number of collisions. For 4G systems, the effect is present but much less pronounced. As noted in section \ref{sec:methods}, we focus on 2-3$^\circ$ inclinations which are more in line with observation.

\subsection{History matters}
\label{sec:history}

Figures \ref{fig:lovis_all_4G} and \ref{fig:lovis_all_3J} show that giant planets can have excursions into the terrestrial zone that are not evident from their final orbits. A giant planet that ventured into the terrestrial zone might later experience a collision, escape the system, or simply move into a wider orbit. Any damage done during the excursion would not be evident from the present-day observable orbits. Figures \ref{fig:lovis_all_4G} and \ref{fig:lovis_all_3J} also show that giant planets can be very destructive \textit{without} venturing into the terrestrial zone. This happens mainly through secular interactions between the outer giant planets and the inner terrestrial planets. Whether rocky planets are destroyed by close encounters, or by long-range secular forces, the implication is the same: \textit{An orbit that is dynamically stable today, may still be empty because it was unstable in the past}. For this reason, it is important to model the history of a planetary system across the dynamical instability, and not just focus on the present-day orbital configuration.

\subsection{Secular evolution}
\label{sec:secular}

Appendix \ref{app:secular} has a brief review of secular theory and forced eccentricity. For this discussion it suffices to say that the value of the forced eccentricities and the width of the secular resonances increase with the eccentricity and mass of the giant planets. In all our runs, the giant planets are initially in circular orbits, but they promptly acquire non-zero eccentricities. The 4Gb runs typically reach an eccentricity of 0.05 after $\sim$650,000 years (median), while the 3J runs take only $\sim$ 1,000 years. As a point of reference, in the present-day solar system Jupiter has an eccentricity of 0.048 and Saturn has an eccentricity of 0.054.

\begin{figure*}
	\includegraphics[width=\columnwidth]{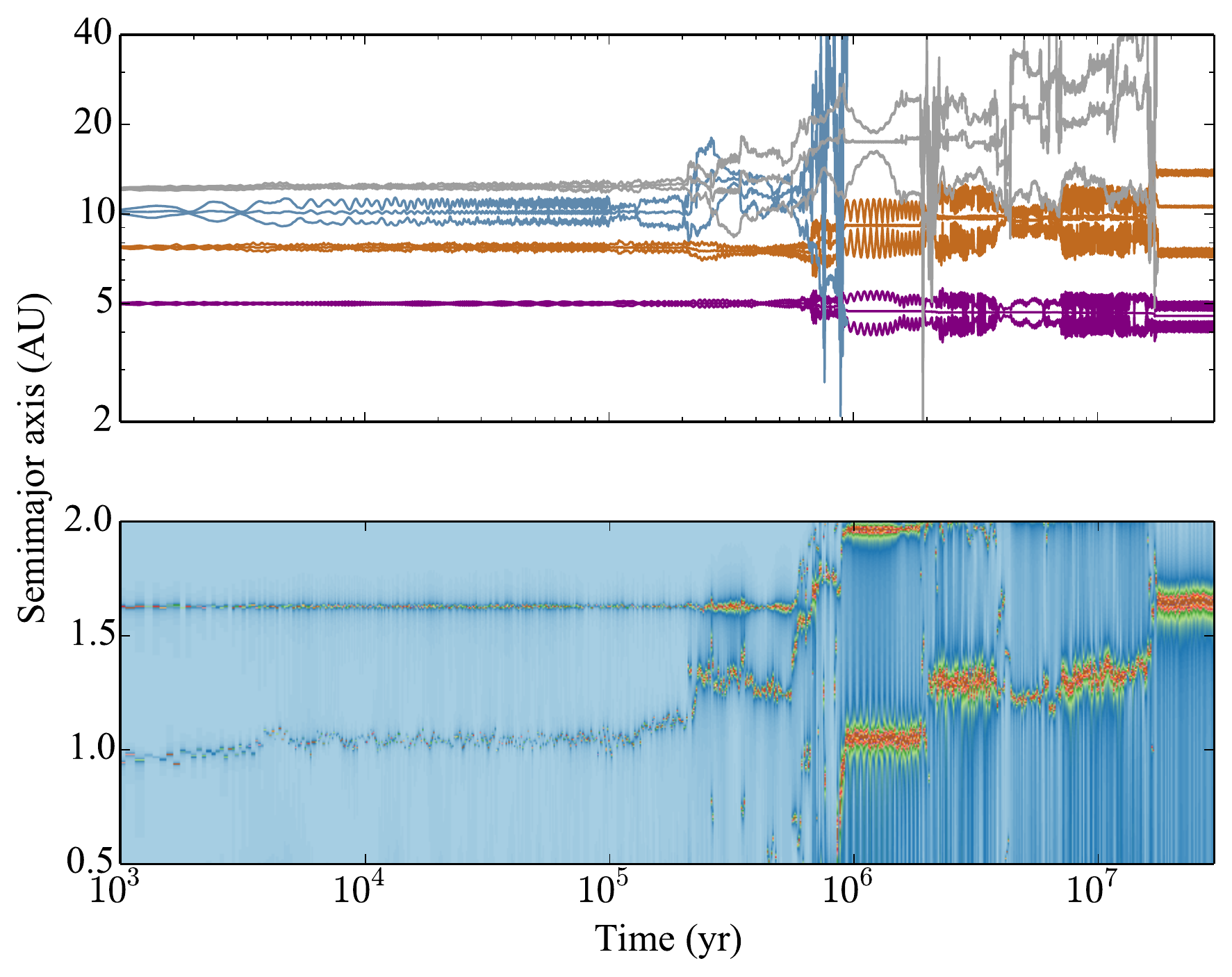}
	\includegraphics[width=\columnwidth]{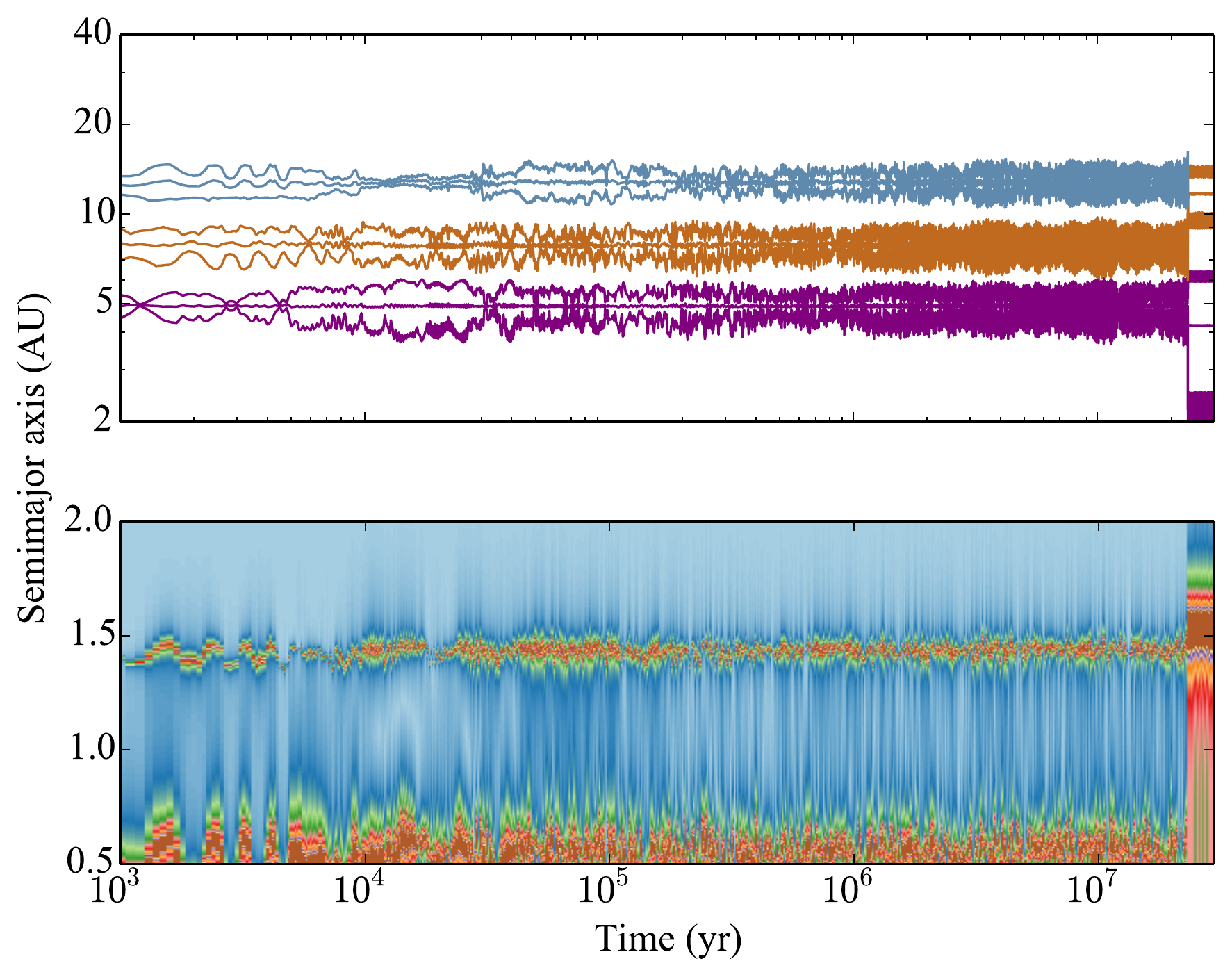}\\
    \includegraphics[width=0.7\columnwidth]{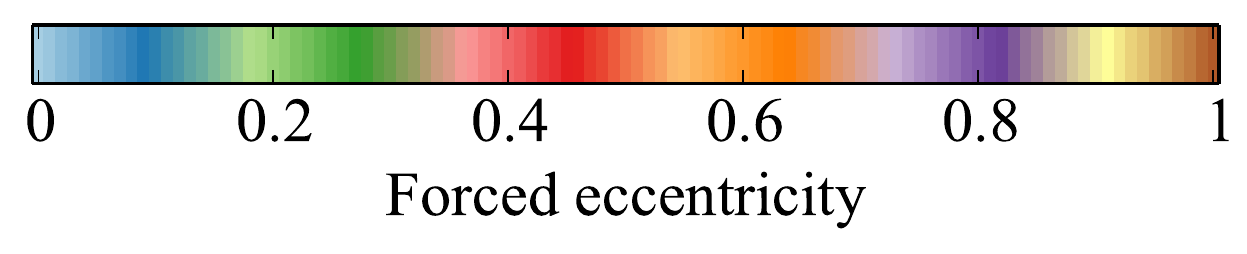}
    \caption{History of one of the 4Gb runs (left) and one of the 3J runs (right). The top plots show the semimajor axes of the giant planets, as well as their periapsis and apoapsis. 4Gb runs become eccentric much later than 3J. The bottom plots shows the forced eccentricities in the terrestrial zone. 4Gb has secular resonances initially a 1 AU and just beyond 1.5 AU, while 3J has them around 0.5 AU and just within 1.5 AU.}
    \label{fig:secular-4Gb-3J}
\end{figure*}

Figure \ref{fig:secular-4Gb-3J} shows the dynamical evolution of an example 4Gb and a 3J system. These two examples were chosen because they are equally destructive to the terrestrial zone -- at the end of the run, both systems are left with 11 surviving particles. This means that we chose one of the more destructive 4Gb systems, and one of the less destructive 3J systems. The bottom half of each plot shows the forced eccentricities produced by the giant planets. Both 4Gb and 3J have two secular resonances in between 0.5 and 2 AU, including one inside the habitable zone; but the secular resonances from 4Gb are narrower and have lower forced eccentricities than in 3J. As the systems evolve, the secular resonances move and grow long before the planets start crossing orbits. In the 4Gb run, the ejection of the two outer giants causes the secular resonance to grow and to sweep through the habitable zone. In the 3J run, the two outer giants experience a collision near the end of the run, leaving a single very wide secular resonance in the habitable zone.

\subsection{Mass and multiplicity of rocky planets}
\label{sec:rocky}

All of the results we have presented so far have relied on test particles as a proxy for terrestrial planets. Test particles allow us to study many terrestrial orbits in parallel, and because $M_\Earth \ll M_\J$, test particles generally do give a good indication of how a terrestrial planet would behave. However, when multiple rocky planets are present, collisions and dynamical interactions can become important. Our next set of simulations has two key goals,

\begin{enumerate}
\item Validate the use of test particles.

\item Explore the effect of rocky planet multiplicity.
\end{enumerate}

\begin{figure}
	\includegraphics[width=\columnwidth]{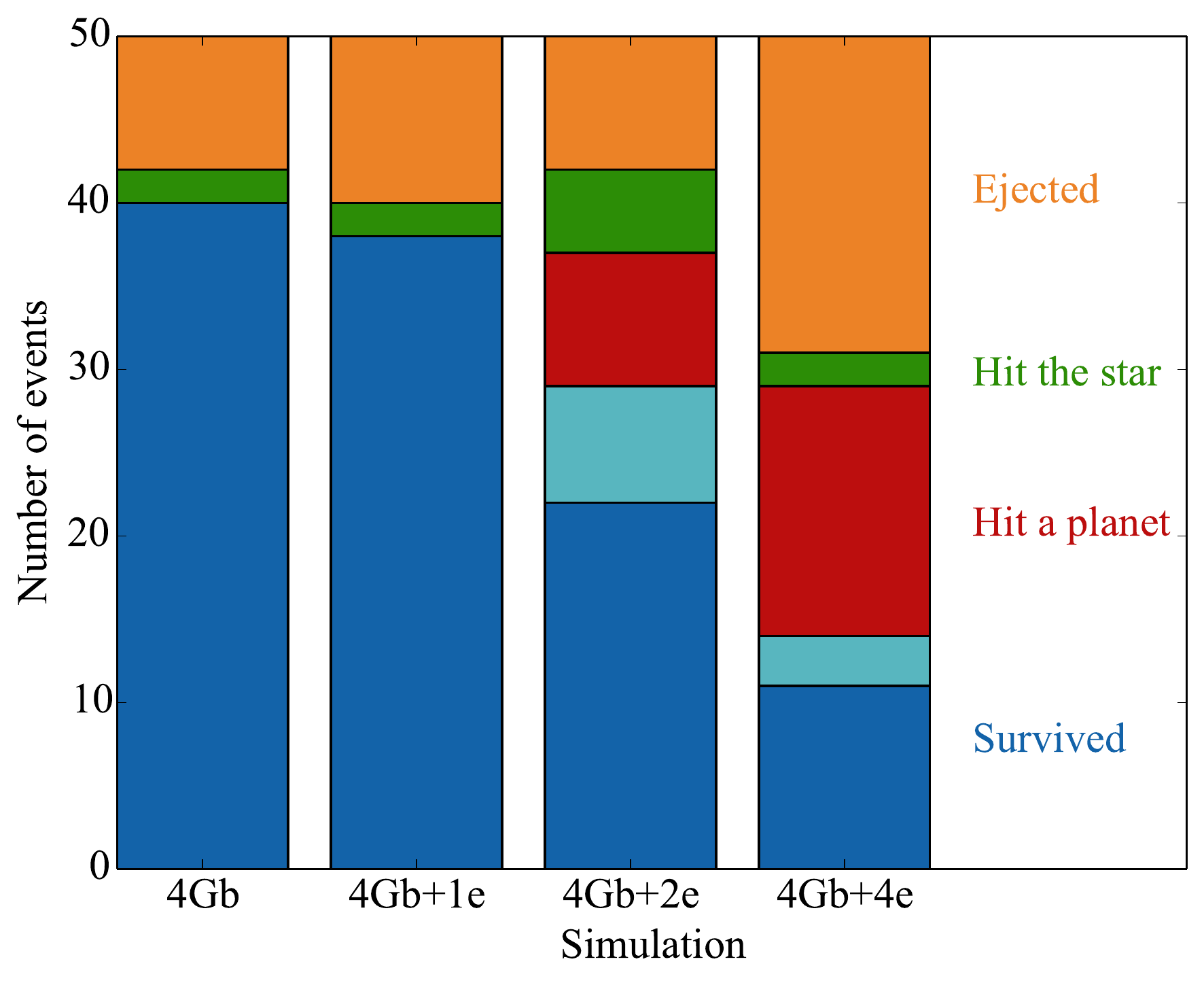}
    \caption{Fate of a planet or test particle at 1 AU. The first column is the 4Gb system with test particles only. 4Gb+1e has a single Earth-mass planet at 1 AU. 4Gb+2e has Earth-mass planets at 0.72 and 1 AU (similar to Venus and Earth). 4Gb+4e has Earth-mass planets at 0.39, 0.72, 1, and 1.52 AU (similar to the orbits of Mercury, Venus, Earth, and Mars). Each system was run 50 times for 30 Myr. The bars show the number of planets (or particles) that were ejected (orange), collided with the Sun (green), or collided with another planet (red), or survived (light and dark blue). The light blue bar is the number of runs where the Earth survives 30 Myr, but is in a crossing orbit with another rocky planet; so it is likely to be destroyed at some point in the future. All planet-planet collisions were with another rocky planet; there were no collisions with giant planets.}
    \label{fig:rocky-survival}
\end{figure}

To this end, we ran simulations of 4Gb systems with the test particles replaced by 1, 2, or 4 rocky planets with $1 M_\Earth$ (see Table~\ref{tab:initial-rocky}). We ran each system 50 times for 30 Myr with random orbits as described in section \ref{sec:methods}. The separation between the rocky planets are $\Delta = 48, 25,$ and 33. In all cases, if the giant planets were not present the rocky planets would be stable for much longer than the 30 Myr simulation time \citep{Chambers_1996}. To verify this, we ran 100 simulations with the four rocky planets in 4Gb+4e (Table~\ref{tab:initial-rocky}) but no giant planets. As expected, after 30 Myr there were zero ejections, collisions, or close encounters. Therefore, any instability observed in runs 4Gb+1e, 4Gb+2e, and 4Gb+4e are ultimately a consequence of the 4Gb giant planets. Figure \ref{fig:rocky-survival} shows that the systems with a single rocky planet (4Gb+1e) behave similar to the test particles. The figure also shows that rocky companions have a damaging effect, largely as a result of direct collisions. This much is expected because rocky planets around 1 AU have escape speeds smaller than their orbital speeds, which favours collisions \citep{Goldreich_2004,Wetherill_1989}. This is can be quantified by the quantity,

\begin{equation}
	\theta^2 = \left(\frac{m_p}{M_\star} \right) \left(\frac{R_p}{a_p} \right)^{-1}
\end{equation}
where $m_p$ and $M_\star$ are the masses of the planet and the star, $R_p$ is the planet radius, and $a_p$ is its semimajor axis. Planets with $\theta \gg 1$ are efficient at ejecting bodies, and those with $\theta < 1$ are more likely to experience collisions \citep{Ford_2008}. But notice that in the system with four rocky planets, the probability that the planet at 1 AU will be ejected or will collide with the Sun is visibly higher. This shows that rocky planets do not behave exactly like test particles, and their dynamical interactions can be consequential. To confirm, we repeated the experiments with 300 runs of 4G and 4G+4e (Table \ref{tab:final-rocky}). We find that rocky companions significantly increase the probability of an ejection, or a collision with the central star.

\begin{table}
	\centering
	\caption{The fate of a rocky planet or test particle at 1 AU in a 4Gb system based on 300 runs with test particles, and 300 runs with three Earth-mass companions (4Gb+4e). There were no rocky-giant planet collisions; all the planet-planet collisions reported are between rocky planets. There was one test particle that collided with a giant planet.}
	\label{tab:final-rocky}
	\begin{tabular}{l|rr}
		\hline
        & Test particles (4Gb) & Rocky companions (4Gb+4e) \\
		\hline
        Ejected      & 17.0\% & 26.7\% \\
        Hit the Sun  &  5.7\% & 11.0\% \\
        Hit a planet &  0.3\% & 31.7\% \\
        Survived     & 77.0\% & 30.7\% \\
		\hline
	\end{tabular}
\end{table}

\begin{figure}
	\includegraphics[width=\columnwidth]{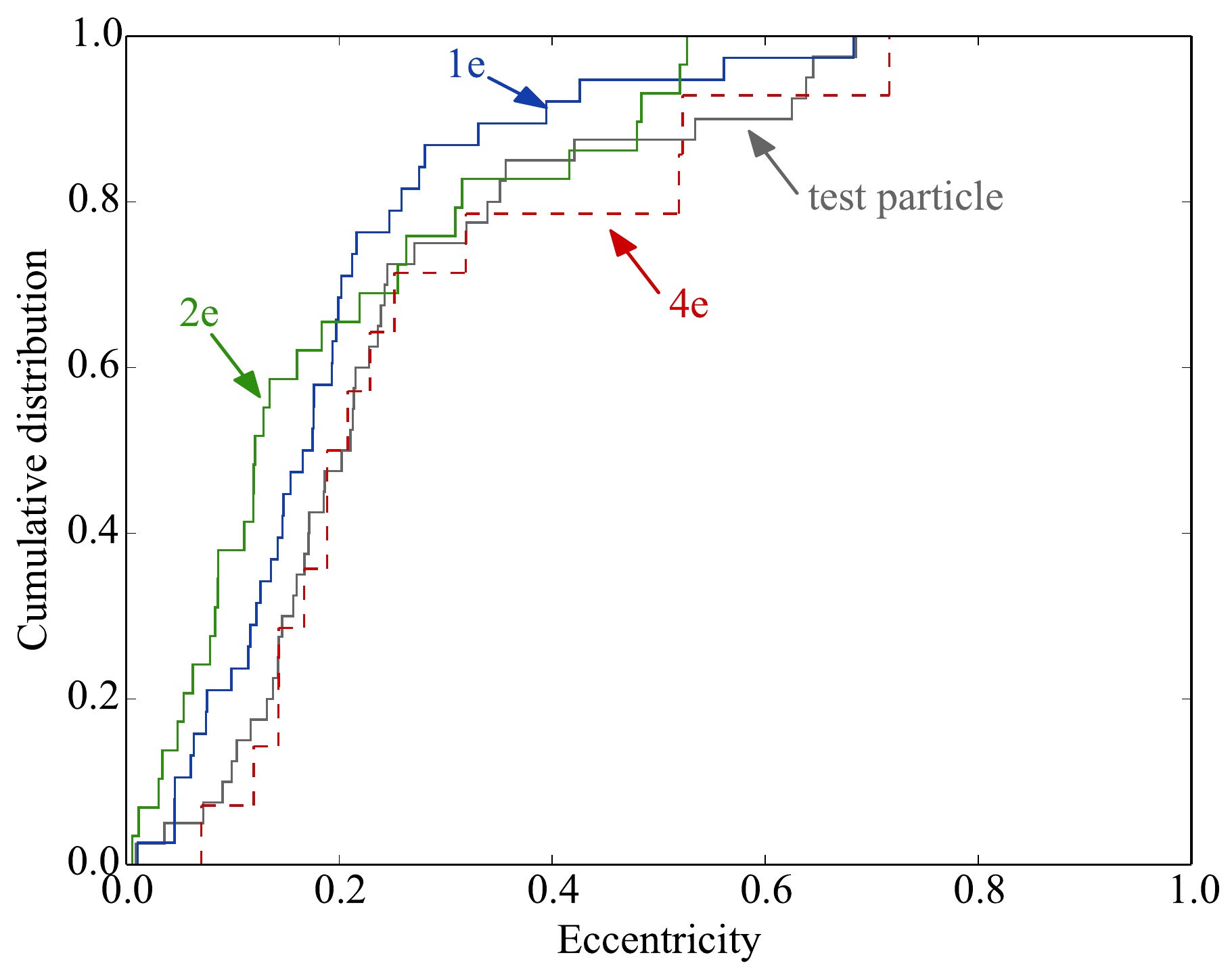}
    \caption{Cumulative distribution of eccentricities of the surviving test particles at 1 AU (grey), and rocky planets at 1 AU for 4Gb+1e (purple), 4Gb+2e (green), and 4Gb+4e (dashed red). Runs with with more than one rocky planet (2e and 4e) do not have systematically different eccentricities than those with a single rocky planet (1e) or a test particle.}
    \label{fig:rocky-eccentricities}
\end{figure}

Figure~\ref{fig:rocky-eccentricities} shows that additional rocky planets do not have a similarly systematic effect on the final eccentricity of the 1 AU planet. The eccentricities look fairly similar. It is possible that a single companion helps dampen the planet's eccentricity, but we leave this investigation for future work.

\subsection{Effect on planet habitability}
\label{sec:habitability}

We chose the 4Gb system to do a more fine-grained measurement of the survival and habitability of terrestrial planets as a function of semimajor axis. In section \ref{sec:rocky} we mentioned an additional set of 300 runs with the 4Gb planets. These runs included 140 test particles, whose semimajor axes are distributed uniformly in log from 0.65 to 3.15 AU. We stop the particles at 3.15 AU because that is the 2:1 resonance with the giant planet at 5 AU. In the solar system, the 2:1 resonance with Jupiter mostly marks the outer edge of the main asteroid belt.

\begin{figure}
	\includegraphics[width=\columnwidth]{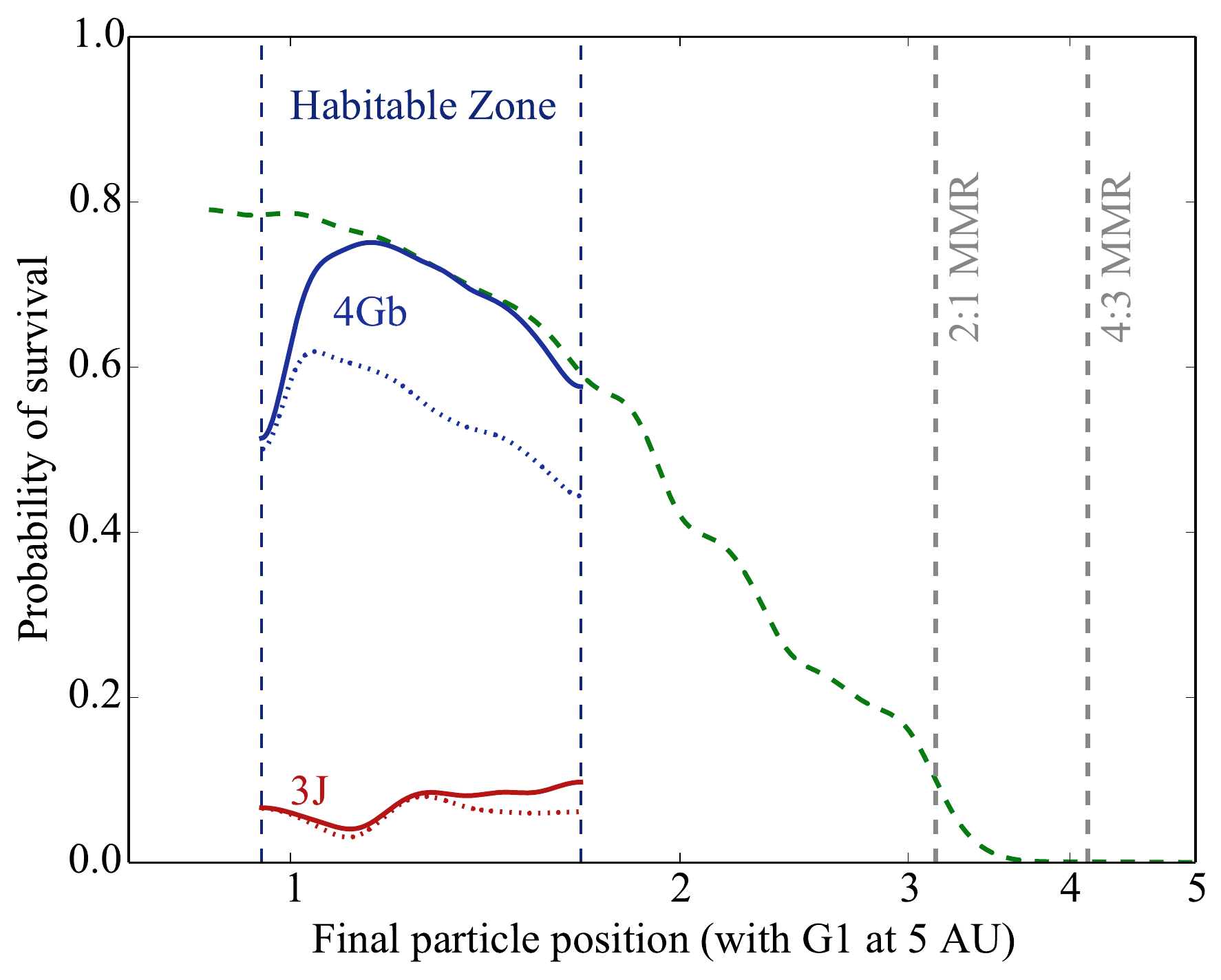}
    \caption{Survival probability of terrestrial planets as a function of semimajor axis for a 4Gb system assuming that the Jupiter-mass planet is at 5 AU (dashed green). In other words, we rescaled the semimajor axes of the particles in each run by a factor $5 \AU / a_{\rm G1}$, where $a_{\rm G1}$ is the semimajor axis of the Jupiter-mass planet at the end of the run. We mark the location of the habitable zone from \citet{Kopparapu_2013b}. The solid blue line shows the probability that a lone habitable planet is \textit{resilient} --- i.e.~retains a mean stellar flux within the habitable zone limits ---. The dotted blue line shows the probability that a habitable planet is resilient and that its mean irradiation changed by less than 10\% at the end of the run (see main text). The red lines show the same results for 3J. The two unresolved 3J runs were excluded from the calculation. The result for 3J is more noisy because of the small number of runs that had survivors. Finally, we show the location of the 2:1 and 4:3 mean motion resonances with the planet at 5 AU. In the solar system, the 2:1 MMR marks the outer edge of the main asteroid belt.}
    \label{fig:habitability}
\end{figure}

Figure \ref{fig:habitability} shows the probability of finding a rocky planet as a function of semimajor axis, renormalized so that the giant planet is fixed at 5 AU. We also mark the conservative habitable zone with the updated values from \citep{Kopparapu_2013b}. We are interested in the probability that a rocky planet that already hosted life would continue to be habitable after the instability. This is difficult to quantify because a planet's obliquity (axial tilt) and eccentricity affect climate in complex ways. For example, eccentricity increases the net stellar flux, and the temperature extremes, while obliquity makes a planet more resistant to global freezing \citep[e.g.][]{Dressing_2010,Abe_2011}. Our approach is to calculate both an optimistic and a pessimistic estimate for continued habitability.

\begin{itemize}
\item The minimum requirement for continued habitability is that the planet never left the habitable zone, and that its mean stellar flux continues to be within the limits of the habitable zone. This is our ``optimistic'' estimate for continued habitability, and is marked as a solid purple line in Fig.~\ref{fig:habitability}.

\item The concept of habitability is is too complex to be captured by a single quantity like mean stellar irradiation. For example, a highly eccentric planet may be sensitive to global freezing at apoapsis, or it might become sterilized at periapsis. Simulations by \citet{Dressing_2010} suggest that a habitable planet can tolerate eccentricities above 0.5 and still have most of its surface habitable over the entire year (figures 6 and 7). For our ``pessimistic'' estimate of continued habitability we add the requirement that the mean irradiation falling on a planet change by no more than 10\% with respect to the mean irradiation that the planet received at the beginning of the simulation. This requirement corresponds to a maximum eccentricity of 0.417 for a planet that did not change semimajor axis. This requirement is marked as a dotted blue line in Fig.~\ref{fig:habitability}.
\end{itemize}

As a point of comparison, we plotted the same two estimates for our 3J runs. This just reiterates the point that 3J systems are more damaging to terrestrial planets. As noted in section \ref{sec:rocky}, if there are other rocky planets present, the true survival rate will be lower.

%
%
\section{Discussion}
\label{sec:discussion}

%
%
\subsection{Habitability as a function of observables}

\begin{figure*}
	\includegraphics[width=1.35\columnwidth]{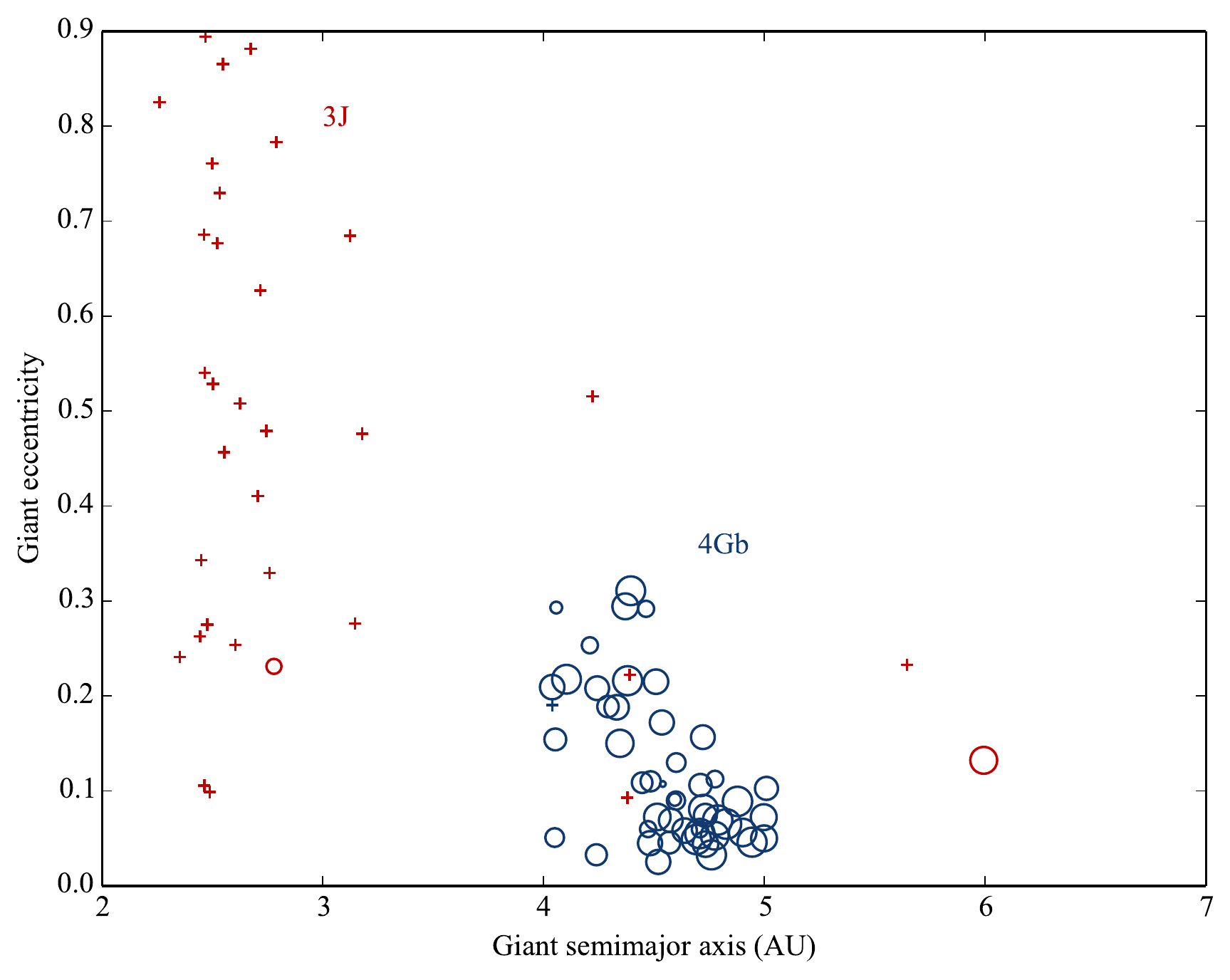}
    \caption{Final eccentricity and semimajor axis of the most massive planet in the 4Gb runs (blue), and the inner giant planet in the 3J runs (red). The two unresolved 3J runs are not shown. The size of each circle is proportional to the percentage of habitable planets that remain habitable, with the largest circle corresponding to 100\%. A plus sign indicates that no habitable planets survived.}
    \label{fig:eJ-aJ}
\end{figure*}

\begin{figure*}
	\includegraphics[height=8.3cm]{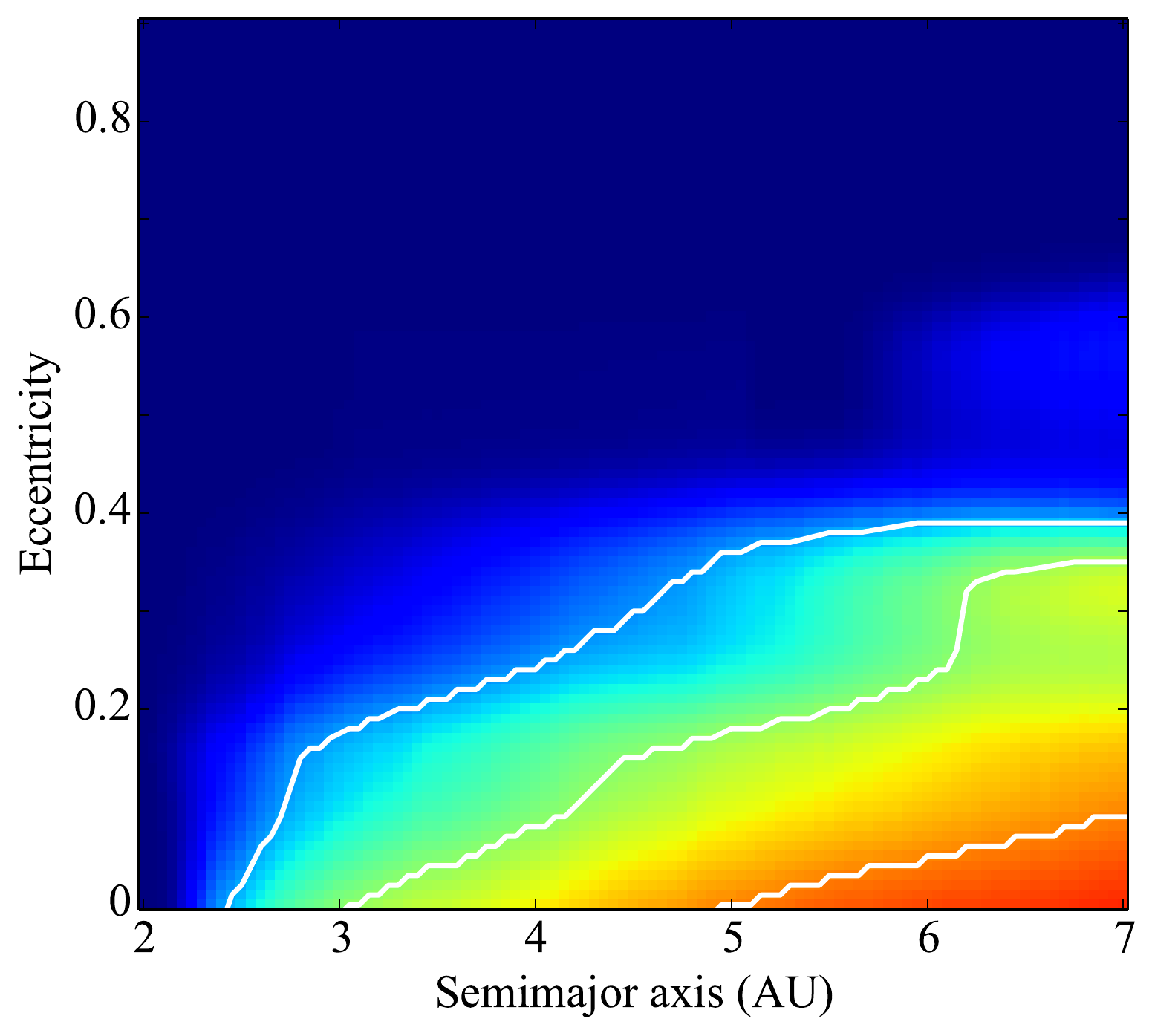}
    \includegraphics[height=8.3cm]{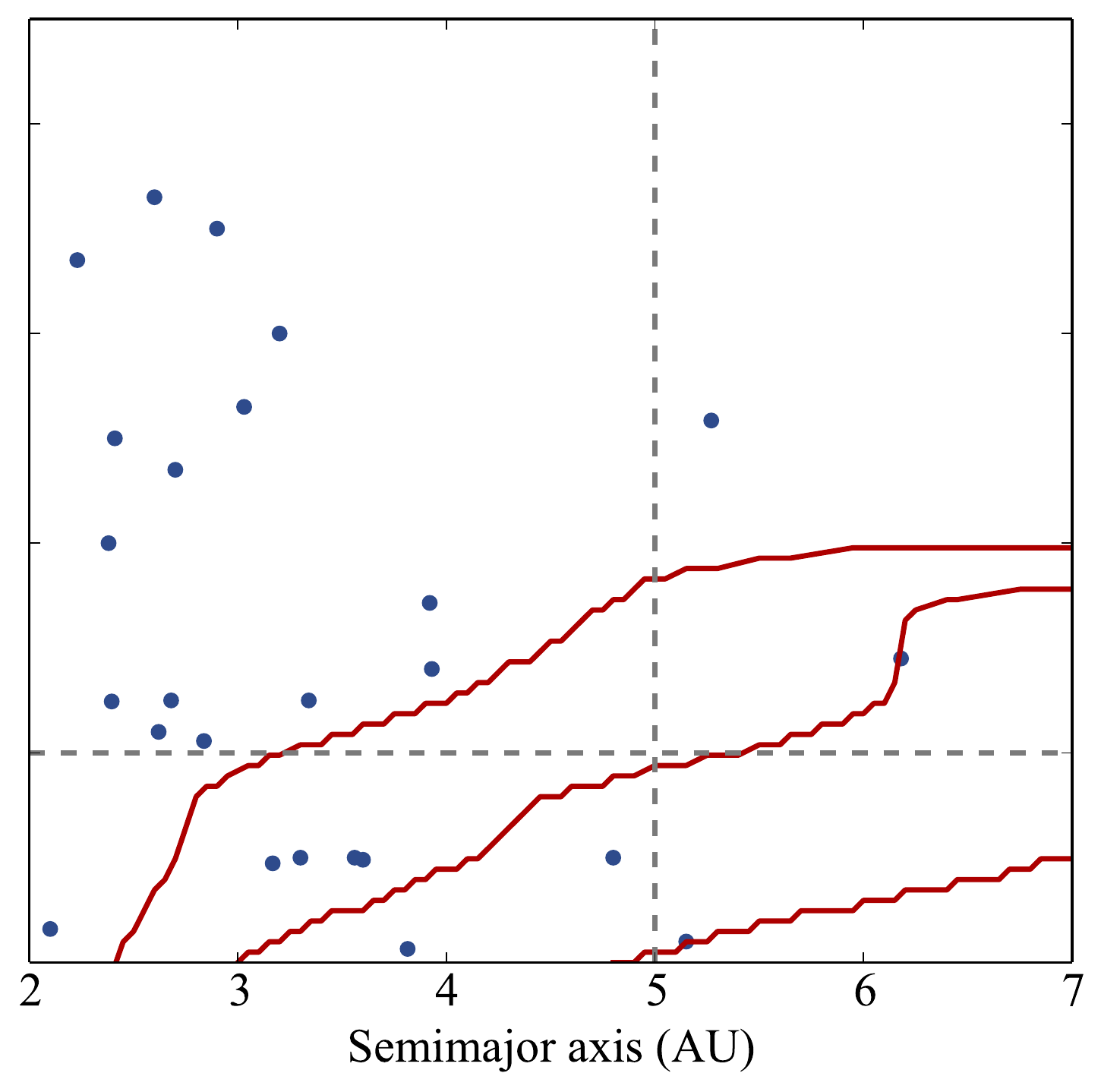}\\
\includegraphics[width=0.7\columnwidth]{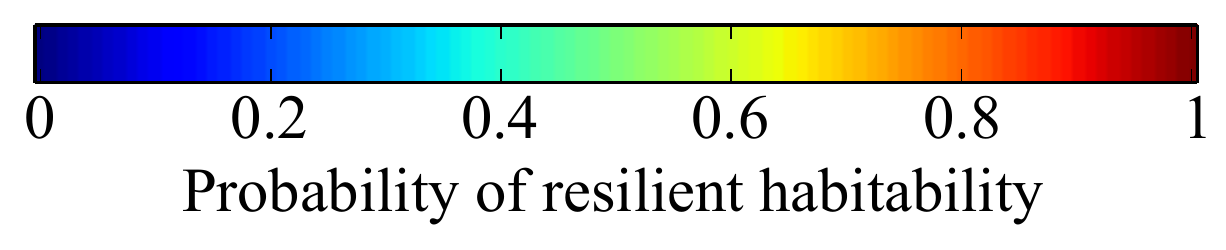}\hspace*{5.18cm}
    \caption{\textit{Left}: Probability that a lone habitable planet has \textit{resilient habitability} --- i.e.~the planet remains habitable after a dynamical instability --- as a function of the present-day semimajor axis and eccentricity of the observable giant planet. The calculation assumes that planet systems are divided between 3J and 4G in a 2:1 ratio, and 4G are split equally between 4Ga, 4Gb, 4Gc. We found that the shape of the plot does not depend strongly on the 3J-to-4G ratio. For each point ($a,e$) we rescale the semimajor axes of our runs and compute the fraction of particles that remain in the habitable zone \citep{Kopparapu_2013a}. We then take a weighted average using Gaussian weights (see main text). The two unresolved 3J runs are excluded from the calculation. The white lines correspond to $p = 0.25, 0.50,$ or 0.75. \textrm{Right}: Currently known exoplanets (blue) with $m > 0.3 M_{\rm J}$ around stars with 0.95 < $M_\star / M_\odot$ < 1.05, along with the $p = 0.25, 0.50, 0.75$ lines (red). For example, for a giant planet with $a = 5$ AU, $e = 0.02$ around a Sun-like star there is a 50\% chance that a habitable planet would be resilient to the instability (dashed grey). The planet next to the $p = 0.75$ line is HD 13931 b (at $a = 5.15$ AU, $e = 0.02$).}
    \label{fig:eJ-aJ-hab}
\end{figure*}

Over the course of a dynamical instability, giant planets can change orbit, collide, merge with the Sun, or escape the system entirely. In section \ref{sec:history} we made the point that an orbit that is dynamically stable today, may still be empty because it was unstable in the past. Despite this complexity, present-day orbits still contain some information about the history of the system. In this section we extrapolate from our simulation results to establish a concrete connection between the present day orbit (semimajor axis and eccentricity) of an observed giant exoplanet, and the likelihood that a habitable planet would have survived to the present day.

Figure \ref{fig:eJ-aJ} shows the final semimajor axes and eccentricities for the most massive giant planet in our 4Gb runs, and the inner giant planet for 3J. For each run we also show the probability that a rocky planet that was initially in the conservative habitable zone \citep{Kopparapu_2013b} was still there at the end of the run. To extrapolate from our simulation results, we present two key ideas:

\begin{enumerate}
\item The first key idea in this section is that orbital dynamics are largely scale free. That is to say, if we take a planetary system and increase all the semimajor axes by (say) 25\%, the dynamical evolution should be the same though on longer time-scales. One caveat is that the likelihood of collision between planets drops with semimajor axis \citep[e.g.][]{Ford_2008}, but as we noted in section \ref{sec:arch}, planet-planet collisions already play a minor role in our simulations. Therefore, we feel that we can justifiably rescale our simulations at least within a narrow range of semimajor axes.

\item The second key idea is that, if we rescale the semimajor axes, a different set of particles will fall in the habitable zone. Taking Fig.~\ref{fig:habitability} as an illustrative example, if we reduce the semimajor axes, a different set of particles will fall in the habitable zone. With the giant effectively moved closer to the habitable zone, the survival rate of habitable planets will drop. Conversely, if we increase the semimajor axes the giant planet will be farther away from the habitable zone and the survival rate of habitable planets will increase.
\end{enumerate}

Let $a$ and $e$ be the semimajor axis and eccentricity of the inner giant planet. Treating $a$ as a free parameter, we can create thousands of virtual systems so that we can probe the $a$ vs $e$ parameter space. For any given value of $a$, we rescale all our runs to put the inner giant at $a$. For each run we compute the fraction of habitable planets that remains habitable at the end of the run $p_{\rm i}(a)$. One important caveat is that the survival of habitable planets also depends on $e$; specifically, more eccentric giants are associated with 3J systems and with lower survival rates. Therefore, we compute a weighted sum across all our runs using Gaussian weights so that the runs where the giant planet eccentricity is closer to $e$ dominate the sum. This gives our final result $p(a,e)$,

\begin{eqnarray}
	p(a,e)   &=& \frac{\sum_{\rm i} \; w_{\rm i}(e) \; p_{\rm i}(a)}{\sum_i \; w_{\rm i}}\label{eqn:p} \\
	w_{\rm i}(e) &=& \exp\left( \frac{(e - e_{\rm i})^2}{2 \; h^2} \right),\label{eqn:weights}
\end{eqnarray}
where the Gaussian weights $w_{\rm i}$ have smoothing length $h = 0.05$. The value $p(a,e)$ estimates the probability $p$ that a habitable planet remained habitable as a function of $a$ and $e$.

\begin{figure}
	\includegraphics[width=\columnwidth]{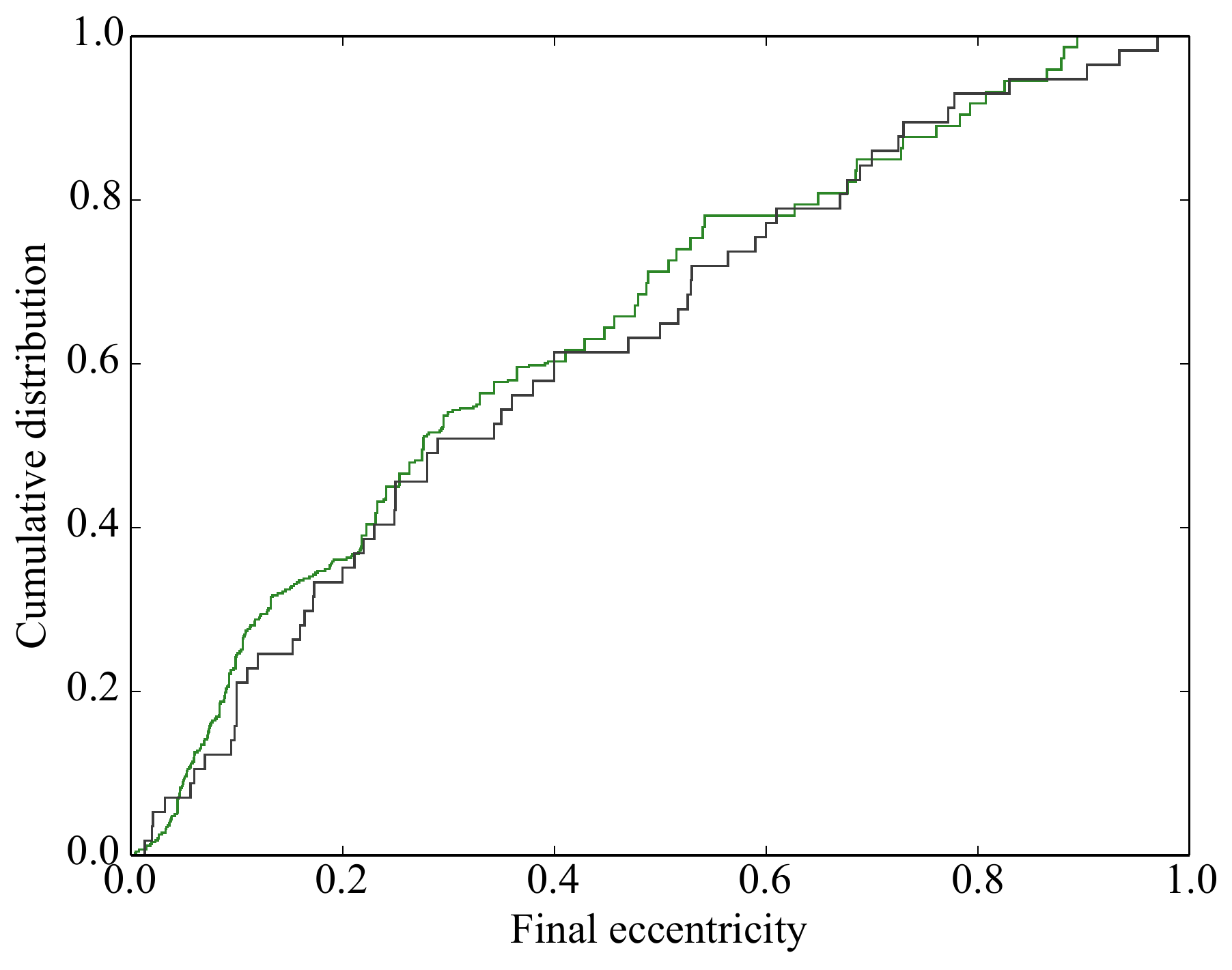}
    \caption{\textit{Black}: Cumulative eccentricity distribution of the observed giant planets plotted in Fig.~\ref{fig:eJ-aJ-hab}. \textit{Green}: Cumulative eccentricity distribution produced by our runs assuming a 2:1 ratio between 3J and 4G systems, and that the 4G are split equally between 4Ga, 4Gb, and 4Gc.}
    \label{fig:cumulative-vs-obs}
\end{figure}

Figure \ref{fig:eJ-aJ-hab} shows the final value of $p(a,e)$ assuming that all giant planet systems are either 3J or 4G, divided in a 2:1 ratio, with the 4G systems divided equally between 4Ga, 4Gb, and 4Gc. In Fig.~\ref{fig:cumulative-vs-obs} we show that that this simple prescription mostly matches the observed distribution of giant planet eccentricities shown in Fig.~\ref{fig:eJ-aJ-hab}. We also tested other ratios and found that the overall shape of the plot is not very sensitive to the 3J-to-4G ratio because 3J systems tend to leave planets in more eccentric orbits than 4G -- the region above $e = 0.3$ is dominated by 3J, while the region below is dominated by 4G.

Figure \ref{fig:eJ-aJ-hab} also marks the points with $p = 0.25, 0.50,$ and 0.75. As a point of reference, giant planets with $e > 0.4$ probably have no habitable companions, and those with $a = 5$ AU, $e = 0.2$ have $p = 0.50$. The plots on the right show that most currently known giant planets around Sun-like stars probably do not have any habitable companions. Currently the best candidate is HD 13931 b ($p \sim 0.75$). As we discover more giant planets beyond 5 AU, the situation will improve.

\subsection{Remark: habitability in stable systems}
\begin{figure}
	\includegraphics[width=\columnwidth]{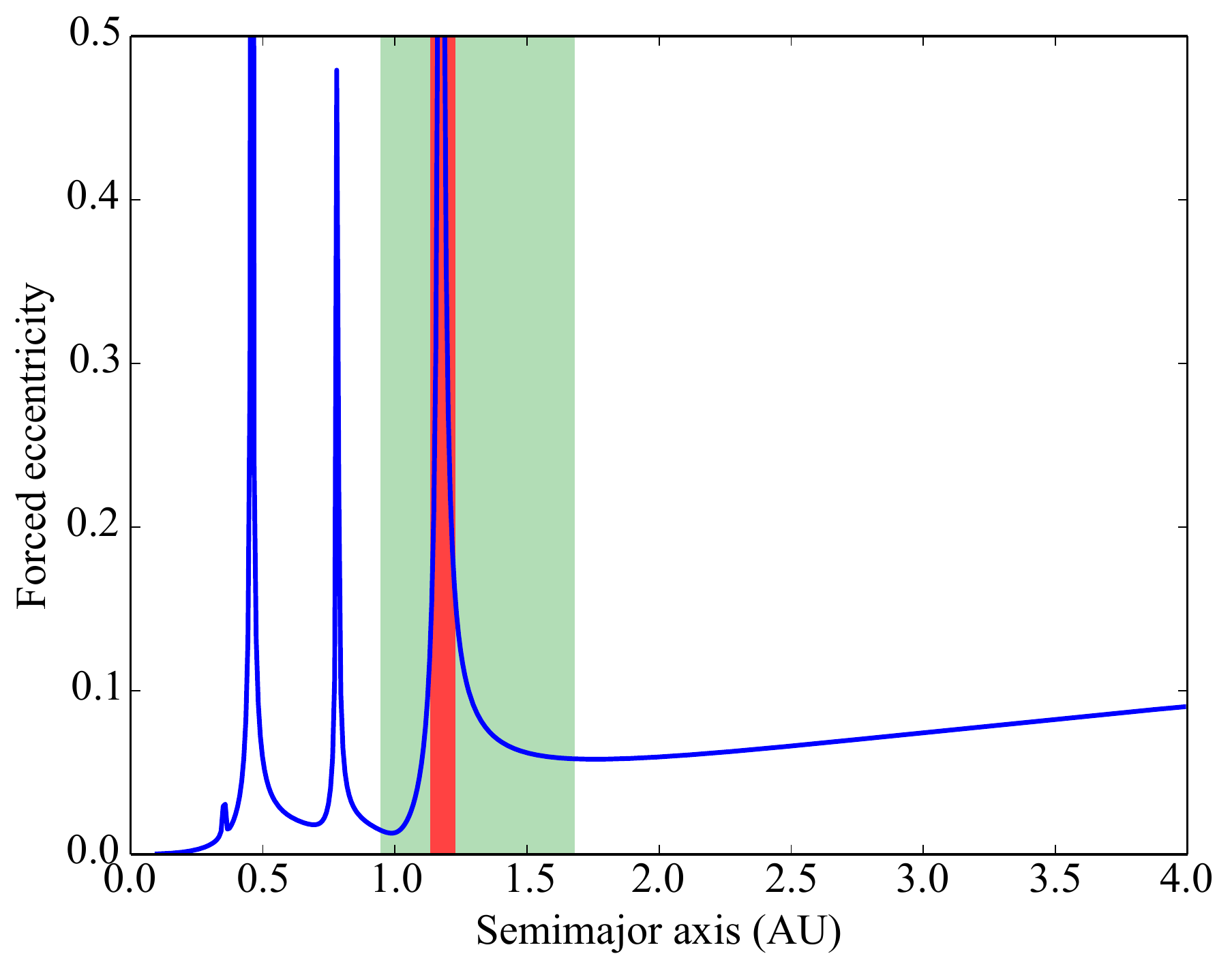}
    \caption{Forced eccentricities for a planetary systems with the same giant planets as 4Gb, but with $\Delta = 10$. All giant planets have $e = 0.1$ and random longitudes of pericentre $\varpi$. This system is probably stable for much longer than the main sequence lifetime of a Sun-like star. The habitable zone is marked in green. There is a secular resonance in the habitable zone. The region where $e_{\rm forced} > 0.15$ is marked in red. If the Earth-Venus system were scaled outward so that the Earth is inside this region, Earth would probably be sterilized by a collision with Venus.}
    \label{fig:secular-hz}
\end{figure}

%
%
Although this paper is about habitability in unstable planetary systems, we would be remiss if we did not point out that Hill instability is not strictly needed for dynamical effects to render a planet uninhabitable. In particular, secular effects can increase the eccentricity of a habitable planet. Figure \ref{fig:secular-hz} is an illustrative example of a stable planetary system that has a secular resonance inside the habitable zone. A habitable planet that finds itself inside a secular resonance will periodically gain high eccentricities. While an the climate of an Earth-like planet at 1 AU may be resilient to eccentricities as high as 0.6, \citep[][figure 4]{Dressing_2010}, a collision between Earth and Venus only requires an eccentricity of 0.27. Whether this effect is at all significant will depend on the semimajor axis and eccentricity distribution of giant planets, which is not currently well understood. We just note that in the solar system the secular resonances are nowhere near the habitable zone, and for planets with Jupiter-like eccentricities, secular resonances are narrow.

%
%
\section{Summary and conclusions}
\label{sec:conclusions}

We have explored the fate of habitable terrestrial planets in planetary systems where the giant planets experience a dynamical instability. The high eccentricities of observed giant exoplanets suggests that dynamical instabilities may have occurred often in planetary systems \citep{Juric_2008}. These instabilities can alter the orbits of terrestrial planets, some times leading to ejections or collisions. In the case of habitable planets, a significant change in the orbital parameters may also take the planet out of the habitable zone. Our work has led to three key results:

\begin{itemize}
\item We find that planetary systems consisting of three Jupiter-mass planets (3J) are extremely destructive to terrestrial planets in these systems, with most runs leading to a complete clearing of the habitable zone (Fig.~\ref{fig:lovis_all_3J}). In contrast, hierarchical systems consisting of four giant planets of unequal masses (4G) are fairly benign to terrestrial planets, with most habitable planets surviving the instability. Within the family of hierarchical (4G) systems, we find that the survival rate of terrestrial planets increases as the giant planets become more hierarchical (Fig.~\ref{fig:survivors}).

\item We establish a concrete link between the present-day orbit of an observed giant exoplanet and the survival of habitable planets. Given the present-day semimajor axis and eccentricity of a giant exoplanet, and provided that the eccentricity was the result of a dynamical instability, we can assign a rough probability that a terrestrial planet in the habitable zone would have survived the instability and remained inside the habitable zone (Fig.~\ref{fig:eJ-aJ-hab}). As a rule of thumb, giant planets with eccentricities higher than 0.4 have experienced strong planet-planet scatterings and are very likely to have originated in a system similar to our 3J systems (Fig.~\ref{fig:eJ-aJ}).

\item Finally, we find that the presence of multiple rocky planets in the system has a harmful effect on the survival of habitable planets. This occurs mainly through physical collisions between rocky planets, but dynamical interactions between rocky planets can also play a role. Depending on the number and the proximity or rocky siblings, the net survival rate can be dramatically reduced. In one set of runs with three rocky siblings, the survival rate for an Earth-like planet dropped to less than half, from 77\% survival rate to 31\% survival rate, with 2/3 of the loss coming from physical collisions with other rocky planets (Fig.~\ref{fig:rocky-survival}; Table \ref{tab:final-rocky}).
\end{itemize}

%
%
\section*{Acknowledgements}

We thank the anonymous referee for many helpful comments and suggestions that allowed us to improve this manuscript. We acknowledge the support from the Knut and Alice Wallenberg Foundation, the Swedish Research Council (grants 2011-3991 and 2014-5775) and the European Research Council Starting Grant 278675-PEBBLE2PLANET that made this work possible. Computer simulations were performed using resources provided by the Swedish National Infrastructure for Computing (SNIC) at the Lunarc Center for Scientific and Technical Computing at Lund University. Some simulation hardware was purchased with grants from the Royal Physiographic Society of Lund.




\bibliographystyle{mnras}
\bibliography{bibliography}

\begin{thebibliography}{}
\makeatletter
\relax
\def\mn@urlcharsother{\let\do\@makeother \do\$\do\&\do\#\do\^\do\_\do\%\do\~}
\def\mn@doi{\begingroup\mn@urlcharsother \@ifnextchar [ {\mn@doi@}
  {\mn@doi@[]}}
\def\mn@doi@[#1]#2{\def\@tempa{#1}\ifx\@tempa\@empty \href
  {http://dx.doi.org/#2} {doi:#2}\else \href {http://dx.doi.org/#2} {#1}\fi
  \endgroup}
\def\mn@eprint#1#2{\mn@eprint@#1:#2::\@nil}
\def\mn@eprint@arXiv#1{\href {http://arxiv.org/abs/#1} {{\tt arXiv:#1}}}
\def\mn@eprint@dblp#1{\href {http://dblp.uni-trier.de/rec/bibtex/#1.xml}
  {dblp:#1}}
\def\mn@eprint@#1:#2:#3:#4\@nil{\def\@tempa {#1}\def\@tempb {#2}\def\@tempc
  {#3}\ifx \@tempc \@empty \let \@tempc \@tempb \let \@tempb \@tempa \fi \ifx
  \@tempb \@empty \def\@tempb {arXiv}\fi \@ifundefined
  {mn@eprint@\@tempb}{\@tempb:\@tempc}{\expandafter \expandafter \csname
  mn@eprint@\@tempb\endcsname \expandafter{\@tempc}}}

\bibitem[\protect\citeauthoryear{{Abe}, {Abe-Ouchi}, {Sleep}  \&
  {Zahnle}}{{Abe} et~al.}{2011}]{Abe_2011}
{Abe} Y.,  {Abe-Ouchi} A.,  {Sleep} N.~H.,   {Zahnle} K.~J.,  2011, \mn@doi
  [Astrobiology] {10.1089/ast.2010.0545}, \href
  {http://adsabs.harvard.edu/abs/2011AsBio..11..443A} {11, 443}

\bibitem[\protect\citeauthoryear{{Armstrong}, {Barnes}, {Domagal-Goldman},
  {Breiner}, {Quinn}  \& {Meadows}}{{Armstrong} et~al.}{2014}]{Armstrong_2014}
{Armstrong} J.~C.,  {Barnes} R.,  {Domagal-Goldman} S.,  {Breiner} J.,  {Quinn}
  T.~R.,   {Meadows} V.~S.,  2014, \mn@doi [Astrobiology]
  {10.1089/ast.2013.1129}, \href
  {http://adsabs.harvard.edu/abs/2014AsBio..14..277A} {14, 277}

\bibitem[\protect\citeauthoryear{{Barnes} \& {Raymond}}{{Barnes} \&
  {Raymond}}{2004}]{Barnes_2004}
{Barnes} R.,  {Raymond} S.~N.,  2004, \mn@doi [\apj] {10.1086/423419}, \href
  {http://adsabs.harvard.edu/abs/2004ApJ...617..569B} {617, 569}

\bibitem[\protect\citeauthoryear{{Batalha} et~al.,}{{Batalha}
  et~al.}{2013}]{Batalha_2013}
{Batalha} N.~M.,  et~al., 2013, \mn@doi [\apjs] {10.1088/0067-0049/204/2/24},
  \href {http://adsabs.harvard.edu/abs/2013ApJS..204...24B} {204, 24}

\bibitem[\protect\citeauthoryear{{Chambers}}{{Chambers}}{1999}]{Chambers_1999}
{Chambers} J.~E.,  1999, \mn@doi [\mnras] {10.1046/j.1365-8711.1999.02379.x},
  \href {http://adsabs.harvard.edu/abs/1999MNRAS.304..793C} {304, 793}

\bibitem[\protect\citeauthoryear{{Chambers} \& {Wetherill}}{{Chambers} \&
  {Wetherill}}{1998}]{Chambers_1998}
{Chambers} J.~E.,  {Wetherill} G.~W.,  1998, \mn@doi [\icarus]
  {10.1006/icar.1998.6007}, \href
  {http://adsabs.harvard.edu/abs/1998Icar..136..304C} {136, 304}

\bibitem[\protect\citeauthoryear{{Chambers}, {Wetherill}  \& {Boss}}{{Chambers}
  et~al.}{1996}]{Chambers_1996}
{Chambers} J.~E.,  {Wetherill} G.~W.,   {Boss} A.~P.,  1996, \mn@doi [\icarus]
  {10.1006/icar.1996.0019}, \href
  {http://adsabs.harvard.edu/abs/1996Icar..119..261C} {119, 261}

\bibitem[\protect\citeauthoryear{{Davies}, {Adams}, {Armitage}, {Chambers},
  {Ford}, {Morbidelli}, {Raymond}  \& {Veras}}{{Davies}
  et~al.}{2014}]{Davies_2014}
{Davies} M.~B.,  {Adams} F.~C.,  {Armitage} P.,  {Chambers} J.,  {Ford} E.,
  {Morbidelli} A.,  {Raymond} S.~N.,   {Veras} D.,  2014, \mn@doi [Protostars
  and Planets VI] {10.2458/azu_uapress_9780816531240-ch034}, \href
  {http://adsabs.harvard.edu/abs/2014prpl.conf..787D} {pp 787--808}

\bibitem[\protect\citeauthoryear{{Dressing}, {Spiegel}, {Scharf}, {Menou}  \&
  {Raymond}}{{Dressing} et~al.}{2010}]{Dressing_2010}
{Dressing} C.~D.,  {Spiegel} D.~S.,  {Scharf} C.~A.,  {Menou} K.,   {Raymond}
  S.~N.,  2010, \mn@doi [\apj] {10.1088/0004-637X/721/2/1295}, \href
  {http://adsabs.harvard.edu/abs/2010ApJ...721.1295D} {721, 1295}

\bibitem[\protect\citeauthoryear{{Faber} \& {Quillen}}{{Faber} \&
  {Quillen}}{2007}]{Faber_2007}
{Faber} P.,  {Quillen} A.~C.,  2007, \mn@doi [\mnras]
  {10.1111/j.1365-2966.2007.12490.x}, \href
  {http://adsabs.harvard.edu/abs/2007MNRAS.382.1823F} {382, 1823}

\bibitem[\protect\citeauthoryear{{Ford} \& {Rasio}}{{Ford} \&
  {Rasio}}{2008}]{Ford_2008}
{Ford} E.~B.,  {Rasio} F.~A.,  2008, \mn@doi [\apj] {10.1086/590926}, \href
  {http://adsabs.harvard.edu/abs/2008ApJ...686..621F} {686, 621}

\bibitem[\protect\citeauthoryear{{Fressin} et~al.,}{{Fressin}
  et~al.}{2013}]{Fressin_2013}
{Fressin} F.,  et~al., 2013, \mn@doi [\apj] {10.1088/0004-637X/766/2/81}, \href
  {http://adsabs.harvard.edu/abs/2013ApJ...766...81F} {766, 81}

\bibitem[\protect\citeauthoryear{{Gladman}}{{Gladman}}{1993}]{Gladman_1993}
{Gladman} B.,  1993, \mn@doi [\icarus] {10.1006/icar.1993.1169}, \href
  {http://adsabs.harvard.edu/abs/1993Icar..106..247G} {106, 247}

\bibitem[\protect\citeauthoryear{{Goldreich}, {Lithwick}  \&
  {Sari}}{{Goldreich} et~al.}{2004}]{Goldreich_2004}
{Goldreich} P.,  {Lithwick} Y.,   {Sari} R.,  2004, \mn@doi [\apj]
  {10.1086/423612}, \href {http://adsabs.harvard.edu/abs/2004ApJ...614..497G}
  {614, 497}

\bibitem[\protect\citeauthoryear{{Johansen}, {Davies}, {Church}  \&
  {Holmelin}}{{Johansen} et~al.}{2012}]{Johansen_2012}
{Johansen} A.,  {Davies} M.~B.,  {Church} R.~P.,   {Holmelin} V.,  2012,
  \mn@doi [\apj] {10.1088/0004-637X/758/1/39}, \href
  {http://adsabs.harvard.edu/abs/2012ApJ...758...39J} {758, 39}

\bibitem[\protect\citeauthoryear{{Johansen}, {Mac Low}, {Lacerda}  \&
  {Bizzarro}}{{Johansen} et~al.}{2015}]{Johansen_2015}
{Johansen} A.,  {Mac Low} M.-M.,  {Lacerda} P.,   {Bizzarro} M.,  2015, \mn@doi
  [Science Advances] {10.1126/sciadv.1500109}, \href
  {http://adsabs.harvard.edu/abs/2015SciA....115109J} {1, 1500109}

\bibitem[\protect\citeauthoryear{{Jones}, {Sleep}  \& {Chambers}}{{Jones}
  et~al.}{2001}]{Jones_2001}
{Jones} B.~W.,  {Sleep} P.~N.,   {Chambers} J.~E.,  2001, \mn@doi [\aap]
  {10.1051/0004-6361:20000078}, \href
  {http://adsabs.harvard.edu/abs/2001A%26A...366..254J} {366, 254}

\bibitem[\protect\citeauthoryear{{Juri{\'c}} \& {Tremaine}}{{Juri{\'c}} \&
  {Tremaine}}{2008}]{Juric_2008}
{Juri{\'c}} M.,  {Tremaine} S.,  2008, \mn@doi [\apj] {10.1086/590047}, \href
  {http://adsabs.harvard.edu/abs/2008ApJ...686..603J} {686, 603}

\bibitem[\protect\citeauthoryear{{Kasting}, {Whitmire}  \&
  {Reynolds}}{{Kasting} et~al.}{1993}]{Kasting_1993}
{Kasting} J.~F.,  {Whitmire} D.~P.,   {Reynolds} R.~T.,  1993, \mn@doi
  [\icarus] {10.1006/icar.1993.1010}, \href
  {http://adsabs.harvard.edu/abs/1993Icar..101..108K} {101, 108}

\bibitem[\protect\citeauthoryear{{Kokubo} \& {Ida}}{{Kokubo} \&
  {Ida}}{1996}]{Kokubo_1996}
{Kokubo} E.,  {Ida} S.,  1996, \mn@doi [\icarus] {10.1006/icar.1996.0148},
  \href {http://adsabs.harvard.edu/abs/1996Icar..123..180K} {123, 180}

\bibitem[\protect\citeauthoryear{{Kopparapu} et~al.,}{{Kopparapu}
  et~al.}{2013a}]{Kopparapu_2013a}
{Kopparapu} R.~K.,  et~al., 2013a, \mn@doi [\apj]
  {10.1088/0004-637X/765/2/131}, \href
  {http://adsabs.harvard.edu/abs/2013ApJ...765..131K} {765, 131}

\bibitem[\protect\citeauthoryear{{Kopparapu} et~al.,}{{Kopparapu}
  et~al.}{2013b}]{Kopparapu_2013b}
{Kopparapu} R.~K.,  et~al., 2013b, \mn@doi [\apj] {10.1088/0004-637X/770/1/82},
  \href {http://adsabs.harvard.edu/abs/2013ApJ...770...82K} {770, 82}

\bibitem[\protect\citeauthoryear{{Levison} \& {Agnor}}{{Levison} \&
  {Agnor}}{2003}]{Levison_2003}
{Levison} H.~F.,  {Agnor} C.,  2003, \mn@doi [\aj] {10.1086/374625}, \href
  {http://adsabs.harvard.edu/abs/2003AJ....125.2692L} {125, 2692}

\bibitem[\protect\citeauthoryear{{Matsumura}, {Ida}  \& {Nagasawa}}{{Matsumura}
  et~al.}{2013}]{Matsumura_2013}
{Matsumura} S.,  {Ida} S.,   {Nagasawa} M.,  2013, \mn@doi [\apj]
  {10.1088/0004-637X/767/2/129}, \href
  {http://adsabs.harvard.edu/abs/2013ApJ...767..129M} {767, 129}

\bibitem[\protect\citeauthoryear{{Mayor} et~al.,}{{Mayor}
  et~al.}{2011}]{Mayor_2011}
{Mayor} M.,  et~al., 2011, preprint, \href
  {http://adsabs.harvard.edu/abs/2011arXiv1109.2497M} {} (\mn@eprint {arXiv}
  {1109.2497})

\bibitem[\protect\citeauthoryear{{Menou} \& {Tabachnik}}{{Menou} \&
  {Tabachnik}}{2003}]{Menou_2003}
{Menou} K.,  {Tabachnik} S.,  2003, \mn@doi [\apj] {10.1086/345359}, \href
  {http://adsabs.harvard.edu/abs/2003ApJ...583..473M} {583, 473}

\bibitem[\protect\citeauthoryear{{Murray} \& {Dermott}}{{Murray} \&
  {Dermott}}{1999}]{Murray_1999}
{Murray} C.~D.,  {Dermott} S.~F.,  1999, {Solar system dynamics}

\bibitem[\protect\citeauthoryear{{Petigura}, {Howard}  \& {Marcy}}{{Petigura}
  et~al.}{2013}]{Petigura_2013}
{Petigura} E.~A.,  {Howard} A.~W.,   {Marcy} G.~W.,  2013, Proceedings of the
  National Academy of Science, \href
  {http://adsabs.harvard.edu/abs/2013PNAS..11019273P} {110, 19273}

\bibitem[\protect\citeauthoryear{{Raymond}, {Quinn}  \& {Lunine}}{{Raymond}
  et~al.}{2006}]{Raymond_2006}
{Raymond} S.~N.,  {Quinn} T.,   {Lunine} J.~I.,  2006, \mn@doi [\icarus]
  {10.1016/j.icarus.2006.03.011}, \href
  {http://adsabs.harvard.edu/abs/2006Icar..183..265R} {183, 265}

\bibitem[\protect\citeauthoryear{{Raymond} et~al.,}{{Raymond}
  et~al.}{2011}]{Raymond_2011}
{Raymond} S.~N.,  et~al., 2011, \mn@doi [\aap] {10.1051/0004-6361/201116456},
  \href {http://adsabs.harvard.edu/abs/2011A%26A...530A..62R} {530, A62}

\bibitem[\protect\citeauthoryear{{Raymond} et~al.,}{{Raymond}
  et~al.}{2012}]{Raymond_2012}
{Raymond} S.~N.,  et~al., 2012, \mn@doi [\aap] {10.1051/0004-6361/201117049},
  \href {http://adsabs.harvard.edu/abs/2012A%26A...541A..11R} {541, A11}

\bibitem[\protect\citeauthoryear{{Rivera} \& {Haghighipour}}{{Rivera} \&
  {Haghighipour}}{2007}]{Rivera_2007}
{Rivera} E.,  {Haghighipour} N.,  2007, \mn@doi [\mnras]
  {10.1111/j.1365-2966.2006.11172.x}, \href
  {http://adsabs.harvard.edu/abs/2007MNRAS.374..599R} {374, 599}

\bibitem[\protect\citeauthoryear{{Selsis}, {Kasting}, {Levrard}, {Paillet},
  {Ribas}  \& {Delfosse}}{{Selsis} et~al.}{2007}]{Selsis_2007}
{Selsis} F.,  {Kasting} J.~F.,  {Levrard} B.,  {Paillet} J.,  {Ribas} I.,
  {Delfosse} X.,  2007, \mn@doi [\aap] {10.1051/0004-6361:20078091}, \href
  {http://adsabs.harvard.edu/abs/2007A%26A...476.1373S} {476, 1373}

\bibitem[\protect\citeauthoryear{{Veras} \& {Armitage}}{{Veras} \&
  {Armitage}}{2005}]{Veras_2005}
{Veras} D.,  {Armitage} P.~J.,  2005, \mn@doi [\apjl] {10.1086/428831}, \href
  {http://adsabs.harvard.edu/abs/2005ApJ...620L.111V} {620, L111}

\bibitem[\protect\citeauthoryear{{Veras} \& {Armitage}}{{Veras} \&
  {Armitage}}{2006}]{Veras_2006}
{Veras} D.,  {Armitage} P.~J.,  2006, \mn@doi [\apj] {10.1086/504582}, \href
  {http://adsabs.harvard.edu/abs/2006ApJ...645.1509V} {645, 1509}

\bibitem[\protect\citeauthoryear{{Wetherill} \& {Stewart}}{{Wetherill} \&
  {Stewart}}{1989}]{Wetherill_1989}
{Wetherill} G.~W.,  {Stewart} G.~R.,  1989, \mn@doi [\icarus]
  {10.1016/0019-1035(89)90093-6}, \href
  {http://adsabs.harvard.edu/abs/1989Icar...77..330W} {77, 330}

\bibitem[\protect\citeauthoryear{{Williams} \& {Pollard}}{{Williams} \&
  {Pollard}}{2002}]{Williams_2002}
{Williams} D.~M.,  {Pollard} D.,  2002, \mn@doi [International Journal of
  Astrobiology] {10.1017/S1473550402001064}, \href
  {http://adsabs.harvard.edu/abs/2002IJAsB...1...61W} {1, 61}

\bibitem[\protect\citeauthoryear{{Wolf} \& {Toon}}{{Wolf} \&
  {Toon}}{2014}]{Wolf_2014}
{Wolf} E.~T.,  {Toon} O.~B.,  2014, \mn@doi [\grl] {10.1002/2013GL058376},
  \href {http://adsabs.harvard.edu/abs/2014GeoRL..41..167W} {41, 167}

\makeatother
\end{thebibliography}



\appendix
\section{Secular perturbation theory}
\label{app:secular}

The N-body problem is nonintegrable for $N > 2$. However, when the system is dominated by a single central body, the orbits of the secondary bodies can be approximated as Keplerian orbits with small perturbations arising from the mutual gravitational attractions between the secondary bodies \citep{Murray_1999}. That is to say, the acceleration on the minor body $j$ is written as,

\begin{equation}
	\ddot{\mathbf r}_j = \nabla_j (U_j + \mathcal{R}_j),
\end{equation}
where $U_j$ is the Keplerian potential and $\mathcal{R}_j$ is known as the disturbing function. As long as the bodies are not near mean motion resonances, the evolution of their orbital parameters can be described by secular perturbation theory. In the case of N planets on coplanar orbits, the disturbing function of planet $j$ can be written as,

\begin{equation}
	\mathcal{R}_j = n_j a_j^2
    	\left[
        \frac{1}{2}A_{jj} e_j^2 + 
        \sum\limits_{k \neq j} A_{jk} e_j e_k \cos(\varpi_j - \varpi_k)
        \right],
\end{equation}
where $n_j, a_j, e_j,$ and $\varpi_j$ are the mean motion, semimajor axis, eccentricity, and longitude of pericentre of planet $j$. The expansion for $A_{jk}$ are given in \citet{Murray_1999}. Conventionally, the values $A_{jk}$ are thought of as the elements of a matrix $\mathbf{A}$ with $N$ eigen values $g_i$. If we now insert test particles into this system, the particle's eccentricity will evolve in time according to,

\begin{eqnarray}
	e \sin \varpi &=& e_{\rm free} \sin(t~\dot{\varpi}_{\rm free} + \beta) + h_0 \\
	e \cos \varpi &=& e_{\rm free} \cos(t~\dot{\varpi}_{\rm free} + \beta) + k_0 \\
    e_{\rm forced} &=& \sqrt{h_0^2 + k_0^2},
\end{eqnarray}
where $e_{\rm free}$ and $e_{\rm forced}$ are known as the free and forced eccentricity. Qualitatively, the particle eccentricity will oscillate around $e_{\rm forced}$ with amplitude $e_{\rm free}$, where $e_{\rm free}$ and $\beta$ are constants set by the boundary conditions. The forced eccentricity is given by,

\begin{eqnarray}
	h_0 &=& - \sum\limits_{i = 1}^N \frac{\nu_i}{\dot{\varpi}_{\rm free} - g_i}
    								\sin(g_i t + \beta_i) \\
	k_0 &=& - \sum\limits_{i = 1}^N \frac{\nu_i}{\dot{\varpi}_{\rm free} - g_i}
    								\cos(g_i t + \beta_i),
\end{eqnarray}
where $\nu_i$ is given in \citet{Murray_1999} and increases with the planet masses; $\beta_i$ is a constant set by the boundary conditions. The important point is that the forced eccentricity diverges when the precession rate of the test particle becomes similar to one of the eigen frequencies of $\mathbf{A}$ (i.e.~$\dot{\varpi}_{\rm free} \approx g_i$). These are known as secular frequencies. In the solar system the $g_6$ eigen frequency is responsible for carving the inner edge of the main asteroid belt.

\begin{figure}
	\includegraphics[width=\columnwidth]{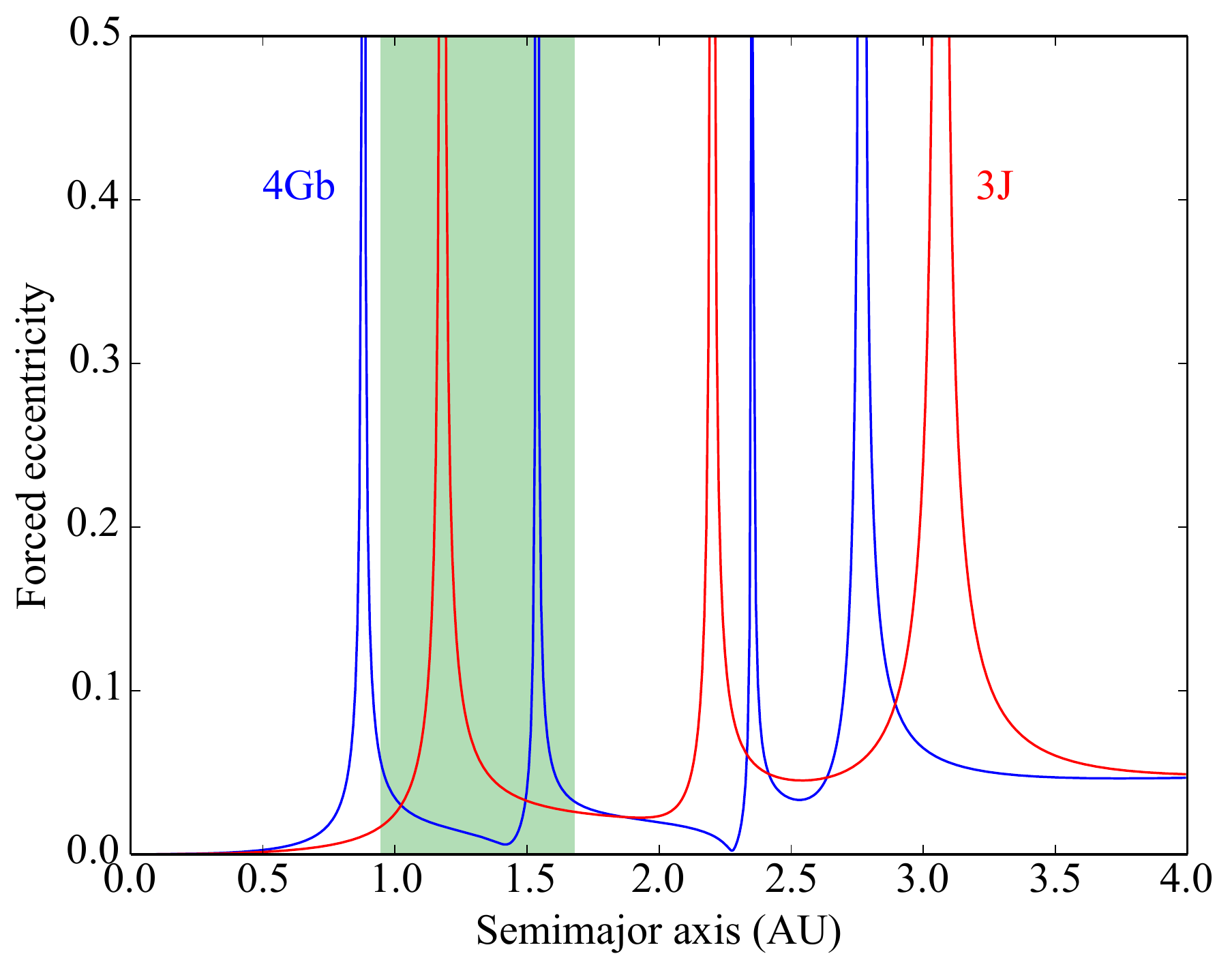}
    \caption{Forced eccentricities of 4Gb and 3J with the giant planet eccentricities all set to $e = 0.05$. 4Gb runs typically reach $e = 0.05$ after $\sim$650 Myr (median), while 3J runs take only $\sim$ 1000 years. See also Fig.~\ref{fig:secular-4Gb-3J}.}
    \label{fig:forced-e}
\end{figure}

Figure \ref{fig:forced-e} shows the forced eccentricities for the 4Gb and 3J systems with the giant planet eccentricities uniformly set to 0.05, and a random choice of $\varpi_j$. As a point of reference, in the solar system Jupiter and Saturn have eccentricities of 0.048 and 0.054 respectively. The discussion so far has focused on a co-planar planet system. For a non-coplanar system, we obtain a similar set of equations involving the inclination $I$ and longitude of ascending node $\Omega$ instead of $e$ and $\varpi$. To first order, the equations for $(e,\varpi)$ are decoupled from those of $(I,\Omega)$.

\bsp	
\label{lastpage}
\end{document}